# AstroECP: towards more practical Electron Channeling Contrast Imaging

Authors

**M. Haroon Qaiser[a], Lukas Berners[b], Robin J. Scales[c], Tianbi Zhang[a], Martin Heller[b], Jiří Dluhoš[d], Sandra Korte-Kerzel[b] and T. Ben Britton[a]***

[a]Department of Materials Engineering, University of British Columbia, Vancouver, British Columbia, Canada

[b]Institute of Physical Metallurgy and Materials Physics, RWTH Aachen University, Germany

[c]Department of Materials, University of Oxford, Oxford, Oxfordshire, United Kingdom

[d] TESCAN Group a.s., Libušina třída 21, Brno, Czech Republic

Correspondence email: ben.britton@ubc.ca

**Synopsis** Here we explore and address many of the major challenges associated with using electron channeling contrast imaging in a scanning electron microscope, with a goal of more easily revealing and characterizing crystalline defects such as dislocations.

**Abstract** Electron channeling contrast imaging (ECCI) is a scanning electron microscopy (SEM) based technique that enables bulk-sample characterization of crystallographic defects (e.g. dislocations, stacking faults, low angle boundaries). Despite its potential, ECCI remains underused for quantitative defect analysis as compared to transmission electron microscope (TEM) based methods. Here, we overcome barriers that limit the use of ECCI including optimizing signal-to-noise contrast, precise determination of the incident beam vector using calibrated, easy to use simulations and experimental selected area electron channeling patterns (SA-ECP). We introduce a systematic ECCI workflow, alongside a new open-source software tool (*AstroECP*), that includes calibration of stage tilting, SA-ECP field of view, and the energy that forms the ECP/ECCI contrast using dynamical simulations. The functionality of this workflow is demonstrated with case studies that include threading dislocations in GaAs and the cross validation of precession based ECCI-contrast, which is otherwise known as electron channeling orientation determination (eCHORD). To assist the reader, we also provide best practice guidelines for ECCI implementation to promote high-resolution defect imaging in the SEM.

**Keywords:** Electron channeling contrast imaging; Selected area electron channeling patterns; eCHORD; pyCHORD; AstroEBSD; dislocations; crystal defects; GaAs epitaxial films

## 1. Introduction

Characterizing crystallographic defects is pivotal for understanding the intricate microstructural behavior of materials and optimizing material performance. Many methods in electron microscopy have been effectively used for analyzing defects such as dislocations and stacking faults in a wide range of materials. Transmission electron microscopy (TEM) has traditionally been the 'go-to' technique for crystallographic defect characterization due to its high spatial and angular resolution, and the collective experience of the materials characterization community. However, TEM lamellae are typically thin (<100 nm), difficult to prepare, and the preparation step can alter the microstructure within the sample, such as the escape of defects to either free surface (due to image forces), and foil bending to relax residual stress (Wu & Schäublin, 2020; Kohnert *et al.*, 2020).

Additionally, the resulting TEM foil typically restricts the observable area to a few localized regions, making quantitative microstructural characterization challenging in many studies. On the other hand, electron backscatter diffraction (EBSD), with its non-destructive nature and ability to map large areas with high spatial resolution, has become a powerful tool to characterize the microstructure of materials at a variety of length scales, revealing grains, grain boundaries, lattice rotations, and elastic strain gradients. Care must be taken when assessing EBSD data with regard to the arrangement of defects, or even individual defects, due to the limited angular resolution for conventional EBSD analysis (Nolze & Winkelmann, 2020). The high angular resolution EBSD (HR-EBSD) method does improve the angular resolution significantly (from 0.5° to ~0.0057°) for misorientation and elastic strain gradient analysis, to unlock characterization of arrays of dislocations including geometrically necessary dislocations (GNDs) (Britton & Hickey, 2018) and, with some assumptions, statistical analysis of the total dislocation density (i.e. statistically stored dislocations, SSDs, and GNDs) (Wilkinson *et al.*, 2014). In the context of the present manuscript and developing of an imaging technique, we recognize that all dislocations are 'geometrically necessary' if they cause a closure failure in the crystal lattice, and the distinction of SSDs and GNDs is only important for EBSD-like measurements, where measurements of the spatial variation in lattice strain are used to estimate the stored dislocation content (i.e. GNDs). Furthermore, for certain samples, it can be possible, although difficult, to measure and analyse the strain field of individual dislocations, e.g. as seen in the HR-transmission Kikuchi diffraction experiments (Yu *et al.*, 2019) and via HR-EBSD (Ernould *et al.*, 2022).

Lately, electron channeling contrast imaging (ECCI) has regained popularity as a promising technique for the direct visualization of dislocations and other crystallographic defects in bulk materials. Although the phenomenon of electron channeling contrast was observed, interpreted and applied several decades ago e.g. (Coates, 1967; Joy *et al.*, 1982; Czernuszka *et al.*, 1990), the technique likely remained relatively obscure due to challenges associated with the implementation of the technique using SEM hardware, and groups interested in deformation analysis in the SEM potentially spent more time with the more rapidly developed 'sibling' – EBSD. However, there is a recent resurgence of interest in ECCI, probably driven by recent developments in SEM hardware such as high-brilliance field-emission guns (also with low energy spread of the electron beam), combined with electron optics that provide optimized beam convergence, high precision eucentric and/or compucentric stages with multiple degrees of freedom, digital control, and large area high sensitivity backscatter electron (BSE) detectors (Gianola *et al.*, 2019). Together these hardware advances provide opportunities to make ECCI more accessible for routine analysis. Additionally, as compared to other techniques, ECCI can be used to scan large areas across the surface of bulk samples and multiple regions of interest, providing improved statistical representativity and a more multiscale approach for studies that require sampling across many grains or locations. It operates at relatively lower electron energies (typically 5–30 keV), which helps reduce damage to the sample and provides information about the defects near to the imaging surface (Crimp, 2006).

In brief, ECCI involves scanning a convergent, and yet nearly parallel, electron beam across the surface of a crystalline sample. When the sample is oriented correctly, electrons can 'channel' with respect to specific lattice directions in the crystal lattice, as partly related to Bragg's law and diffraction-like conditions for the chosen incoming electron beam vector [uvw] with regard to the orientation of the crystal. In the absence of crystalline defects within a region of common misorientation, there would be a uniform signal with no contrast as the amount of electron channeling is the same for every sampled point. If there is a crystallographic defect, then the channeling condition of the electron beam varies around the defect, and so the defect appears as a region of contrast within the micrograph associated with the type of defect and how this perturbs the lattice, and in particular the [uvw] vector of the incident beam with respect to the matrix and defect. As an example, it is possible to image threading defects in crystals, which typically appear as black and white spots in the ECCI micrograph (Zaefferer & Elhami, 2014; Crimp, 2006). As the dislocation changes the local Bragg condition in the lattice, this can modulate the BSE yield if an appropriate incident beam vector [uvw] is selected for defect imaging (Hiller *et al.*, 2023). The spatial variation of contrast depends on several factors, including the incident beam vector [uvw], the electron wavelength ($\lambda$), the Burgers vector ($\vec{b}$) and line direction ($\vec{l}$) of the dislocation under observation, any surface relaxation due to the image forces, and the location of the dislocation with respect to the free surface.

A brief survey of the literature highlights that ECCI has been successfully applied to a wide range of materials to help answer a variety of questions in materials science, such as metals and alloys (Gutierrez-Urrutia *et al.*, 2009*a*; Ng *et al.*, 1998; Britton, Goran *et al.*, 2018; Goldbaum *et al.*, 2015; Ahmed *et al.*, 2001), semiconductors like GaAs/Ge and SiGe/Si (Mangum *et al.*, 2022; Wilkinson *et al.*, 1993), geological minerals and ceramics (Miyajima *et al.*, 2019, 2018; Martin *et al.*, 2021; Rogowitz *et al.*, 2018), and thin films (Trager-Cowan *et al.*, 2006; Naresh-Kumar *et al.*, 2016). The last three decades have seen significant advancements in the development and use of ECCI for specific applications. Researchers have applied ECCI to identify dislocations and determine their Burgers vectors in single crystals (e.g., GaN), polycrystals (e.g., 2 % Si electrical steels), commercially pure titanium, and $UO_2$ ceramics (Mansour *et al.*, 2014; Picard *et al.*, 2009; Mansour *et al.*, 2019; Han *et al.*, 2018). Researchers have also used it to observe dislocation motion in real time under mechanical loading and other *in situ* studies (Nakafuji *et al.*, 2020; Ben Haj Slama *et al.*, 2021, 2019), and for a correlative approach to link EBSD data with ECCI (Gutierrez-Urrutia *et al.*, 2013; Dunlap *et al.*, 2018; Dorri *et al.*, 2016; Mandal *et al.*, 2025). Additionally, simulations and contrast models have been developed to better understand how contrast is formed in ECCI micrographs (Pascal *et al.*, 2018; Twigg *et al.*, 2010; Picard *et al.*, 2014; Kriaa *et al.*, 2019, 2021)

Even with these beautiful case studies that showcase the potential of ECCI to aid understanding of materials, and given the ready availability of scanning electron microscopes for most research labs, the use of ECCI remains relatively limited. It is likely that this is related to the barriers for a new facility to start performing ECCI-based materials characterization. One major challenge is that ECCI is highly sensitive to variations in the crystal (i.e. the precise reason why the technique is useful for defect analysis), and as such it is also extremely sensitive to subtle misalignments. To reveal defects confidently using ECCI, precise sample to beam alignment, with a requirement of [uvw] zone orientation accuracy better than 0.1° to specific features within the electron channeling pattern (ECP), is essential for obtaining high-quality ECCI micrographs (Mansour *et al.*, 2014; Kriaa *et al.*, 2017; Vystavěl *et al.*, 2019). Furthermore, to access high contrast conditions it is important to be able to prepare the sample surface and present an appropriate [uvw] crystallographic direction along the incident optic axis of the microscope, and this requires access to rotation and tilting of the sample within the chamber combined with accurate knowledge of the updated [uvw] vector. Generally, ECCI is challenging because the user must have access to a detailed knowledge of how to navigate the channeling conditions within each grain of their sample to realize reliable defect contrast. This knowledge is materials and sample dependent, for example in thin films the orientation and the types of defects in the film are often well known *a priori* (e.g. threading dislocations in a film that is grown along the <c> axis of the cell). Extending ECCI-based defect imaging to more general experiments can be time consuming, frustrating and challenging, and perhaps this has limited the wide-spread adoption of the technique (e.g. as compared to the application of EBSD). These limitations, along with the need for high-end SEM systems with sensitive, well positioned and large area BSE detector, has likely restricted ECCI's widespread use.

With improvements to microscopes, especially through digital control and a better understanding of the electron column optics, many electron microscopes can now operate in more than one mode:

- (Mode 1) imaging - the electron beam is focussed on the sample, and the scan coils deflect this highly focussed and (near-)parallel electron beam across the surface of the sample. To raster the beam and form an image, the scan coils tilt the beam across the surface of the sample. However, at high magnifications[1], the tilt angle between points in the raster is considered negligible such that the incident beam vector, [uvw], is common for the whole area scanned and typically perpendicular to the surface of the micrograph.

- (Mode 2) channeling - the scan coils and lens configuration are adjusted to enable the beam to strike the same area within the sample, but each point in the micrograph is now systematically adjusted to vary the incident beam direction with respect to the sample such that a 'selected area electron channeling pattern' (SA-ECP, or SACP)[2] can be formed. These patterns can contain crystallographic information in the form of bands of raised intensity, i.e. Kikuchi patterns, which look similar to highly magnified EBSD-patterns.

In mode 1, crystallographic contrast may be visible, even in the absence of any related changes in chemistry or local topography, as the signal collected can be modulated by any variation in electron channeling as the electrons enter and escape the sample, e.g. due a change in orientation of the crystal, or local perturbations in the lattice due defects like dislocations. However, correct interpretation of this contrast requires precise knowledge of the incident beam direction with regards the crystal orientation.

To effectively perform ECCI-based analysis of crystallographic features using SA-ECP, we require accurate understanding of the (mode 2) SA-ECP collection parameters with respect to the incident beam direction used to collect each (mode 1) imaging micrograph. Firstly, it is useful if we can align the (mode 2) SA-ECP to have the central point equivalent to the direct beam used in the (mode 1) imaging and be aware of any chances in column configurations or design that can cause any misalignment. Secondly, we desire easy navigation of the SA-ECP in order to select (mode 1) imaging conditions, which requires a stage that can achieve precise and repeatable stage tilts and/or stage rotations (note these can be validated through SA-ECP collection and analysis).

A limitation of SA-ECP informed ECCI is the cost of ECP collection (as high contrast ECPs typically require long dwell times), the need to have a large area and high contrast detector, and the microscope must be able to form the ECP (i.e. mode 2). Furthermore, aberrations in the microscope may limit the size of the selected area, which can restrict the size of grains that can be analysed and also cause distortions in the ECP if the grains contain significant lattice rotation gradients. ECPs often are collected from a relatively small angular field of view, which may make identification of the specific zone axis challenging.

---

[1] At low magnifications, many electron microscopes deflect the beam at significant angle with the surface of the crystal, and this can result in changes in the direction [uvw] with respect to the crystal. Thus, at very low magnifications, wide angle electron (rocking) channeling patterns can be observed if the sample is a single crystal of uniform orientation

[2] In this manuscript we use the terminology 'selected area electron channeling pattern', SA-ECP, as we recognize that channeling patterns can be obtained from wide-angle 'rocking', and also that other forms of channeling patterns exist in the electron microscopy field, such as ion channeling patterns.

The literature does provide a potential route to address this challenge, through a combination of EBSD and ECCI, i.e. using EBSD to provide knowledge of the ECCI conditions. In one route, Zaefferer and Gutierrez-Urrutia introduce a method they call "ECCI under controlled diffraction conditions" or cECCI (Gutierrez-Urrutia *et al.*, 2009*b*; Zaefferer & Elhami, 2014), which makes use of high-tilt EBSD scans (typically 70°) to determine the crystallographic orientation, followed by simulations in software such as *TOCA* (*tools for orientation and crystallographic analysis*) that predict the necessary stage movements to achieve optimal channeling conditions at low tilts (Zaefferer, 2000, 2003). As a reminder, EBSD is now a very commonly used microscopy tool, and regular use has a precision of ~2° for absolute orientation determination (Nolze & Winkelmann, 2020), 0.5° to 0.1° for misorientations via conventional EBSD analysis (Humphreys, 1999; Wilkinson, 2001; Wilkinson *et al.*, 2006), and misorientations of 0.05° to 0.01° via pattern matching methods (Winkelmann *et al.*, 2020; Brough *et al.*, 2006; Wilkinson *et al.*, 2009). This means that EBSD-informed-ECP/ECCI analysis could be attractive to help guide the selection of a [uvw] axis within the ECP, especially if the ECP is captured across a narrow rocking angle. This approach has been applied in numerous studies over the years in variety of materials (Gutierrez-Urrutia *et al.*, 2010; Zhang *et al.*, 2015; Gutierrez-Urrutia & Raabe, 2012; Gutierrez-Urrutia *et al.*, 2013). However, the reported angular accuracy in such cECCI method is ~0.5° owing to the uncertainties in EBSD indexing and limitations in SEM stage accuracy, which in turn may lead to discrepancies between the predicted and realized channeling conditions (Gammer & An, 2022), and these will be explored as part of the present work. We also note that there are other methods that can link EBSD to ECCI, including the emerging potential use of the RKD-based EBSD geometry (Veghte *et al.*, 2024). During peer review, we were also made aware of an emerging python-based tool, OpenECCI, which includes ECP reprojection of EMSoft simulations and analysis, built using data collected from a Themo Fisher system (Xu *et al.*, 2024*a*,*b*).

For our lab, to assist in making ECCI more accessible and building upon our experiences with EBSD analysis, we have been motivated to develop a complementary software tool. This tool, called *AstroECP,* is introduced and validated in this manuscript and is a *MATLAB*-based GUI created as an extension to the *AstroEBSD* framework specifically designed to assist in stage navigation and pattern indexing for ECCI. The original code also incorporates various functions from open-source *AstroEBSD* (Britton, Tong *et al.*, 2018) and *MTEX* (Bachmann *et al.*, 2010). *AstroECP* reads the dynamically calculated Kikuchi reference patterns generated from pattern simulation software that include: (a) *MapSweeper* EBSD dynamical patterns; (b) higher quality Bloch Wave Kikuchi Diffraction (BWKD) patterns approach which is the model also used by the Oxford Instruments *MapSweeper* product but generated with a custom Python front end); (c) the open-source software *EMsoft* (Singh *et al.*, 2017); and (d) Bruker *DynamicS*. The software then reprojects the channeling pattern based upon the crystal orientation and ECP conditions. Manipulation of the simulation can be performed through virtual movement of the stage and microscope to inform movement of the sample within the SEM. This helps to effectively orient the crystal for the development of crystallographic contrast with a specific [uvw] direction along the optic axis of the SEM and provide imaging conditions that are suitable for the collection of ECCI micrographs.

Additionally, the reprojection algorithms and systematic collection of SA-ECPs provide a unified approach to link together the new method for crystalline orientation mapping in a conventional SEM called electron channeling orientation determination (eCHORD) using only a standard goniometer and BSE detector, as developed originally by Lafond and colleagues (Lafond *et al.*, 2018, 2020). In the eCHORD method, the channeling contrast is recorded as a function of sample rotation about its tilted normal: at each angular step, the BSE intensity varies according to the deviation of the incident beam from exact Bragg conditions, producing a unique intensity profile that can be matched to simulated ECPs to recover the local crystal orientation with an angular resolution better than 1°. As a note, the eCHORD approach is related to the method of rotational ECCI (R-ECCI) reported elsewhere in the literature (L'hôte *et al.*, 2019).

Overall, this manuscript presents a carefully calibrated and demonstrated experimental workflow and software approach, including presentation of *AstroECP*. Initially, the dynamical calculations of the Kikuchi diffraction patterns are precisely compared to establish an appropriate simulation energy to reproduce relevant contrast within the SA-ECP and ECCI micrographs. Next, systematic ECP and ECCI imaging experiments are performed to provide calibrated understanding of the microscope (including stage tilt/rotation axes, selected area convergence angles etc.) to realize rapid navigation of an experimental SA-ECP. These experiments enabled parallel development of the *AstroECP* tool that was used for the analyses including systematic experiments: (i) measurement of tilt and rotation series, and precise pattern matching of experimental and simulated ECPs, to map out stage movements within the microscope; (ii) measurement of the variation in rocking angle with working distance to establish and assist in optimization of the SA-ECP rocking angle; (iii) confirmation of the incident beam vector for ECCI and how that relates to the SA-ECP. In the experiments, the precision of the method is explored to provide additional understanding of ECP-related experiments. Demonstration of the workflow has been provided via a systematic analysis of the contrast associated with threading dislocations in GaAs photonic materials. To provide some context of this approach, the workflow has also been supported by consideration and evaluation of prior art in ECCI-based defect imaging, including an experiment that closely replicates what could be done in a proxy to a cECCI approach, as well as comparison with the precession-BSE based eCHORD methods. Finally, to broaden the reach of the ECCI approach, this paper provides some best practice guidelines with a view to opening up the technique to an even wider user base.

## 2. Materials and methods

### 2.1. Samples

A high-purity, semiconductor grade single crystal of [001] silicon measuring approximately $20 \times 30 \times 1$ mm$^3$ was used to collect SA-ECPs at various stage tilt and rotation configurations to assess uncertainties in stage alignment and direct beam (Section 3). The same crystal was also employed in the precession series to correlate BSE contrast with SA-ECPs (Section 6).

Additionally, three smaller pieces of [001] silicon were sectioned from a standard 1 mm thick single crystal using a diamond scribe and wafer-cleaving pliers. These pieces were co-mounted on a standard $\phi$12.7 mm aluminum stub to make a *pseudo* tri-crystal sample, with deliberate in-plane rotations about the [001] axis to introduce some misorientations between the three pieces, and slight out of plane misorientations associated with how well the samples were affixed to the stub. This configuration was used to correlate misorientations in their SA-ECPs and electron backscatter patterns (EBSPs), as described in Section 5.

As a final sample, an epitaxially grown GaAs layered sample was used for ECCI to study surface threading dislocations (Section 7). The GaAs top layer was approximately 100 nm thick and was followed by indium- and phosphorus-doped variants. The difference in crystal structure and lattice parameters of GaAs (zinc blende lattice) and Ge (diamond cubic lattice) gives rise to misfit dislocations in these epitaxial layers. This application is part of a wider study with our collaborators, as these materials are currently being investigated for the development of vertical-cavity surface-emitting laser (VCSEL) structures (for more information, see reference (Wan *et al.*, 2024)).

### 2.2. Methods

The electron-microscopy experiments were carried out in a TESCAN AMBER-X plasma-FIB field-emission scanning electron microscope (pFIB-FESEM) operated at 20 keV and a base pressure $< 5\times10^{-4}$ Pa. A large-area four-quadrant backscattered-electron detector (4Q-BSE) that can be inserted and positioned below the pole piece, collected both ECCI micrographs and SA-ECP used in this study.

Pattern indexing, geometry (i.e. pattern-center) determination and kinematic-band labelling were performed in the open-source *AstroECP* that is a graphical user interface (GUI) developed in *MATLAB*, based on the approaches and functions used in *AstroEBSD* (Britton, Tong *et al.*, 2018) and *MTEX* v5.10.2 (Bachmann *et al.*, 2010; Hielscher *et al.*, 2019) and this is introduced in more detail in Section 4. For the stage calibration and precession series experiments, which required capturing a large number of BSE images and SA-ECP at various stage positions, an automated SEM control approach was implemented using the TESCAN *SharkSEM* v3.2.7, TESCAN Essence Emulator, and a Python-based application programming interface (API).

Post processing of ECPs, where indicated, includes a low frequency bandpass filter to suppress the background and increase contrast of high spatial frequency features inside zone axis. For the beam current experiment, a 2D-FFT Hann windowing was applied to suppress ringing artifacts within the evaluation process to measure the signal to noise ratio using a radial averaging method in *MATLAB*.

EBSD data were recorded using a Symmetry S2 detector (Oxford Instruments, UK) with the sample loaded into a pre-tilted holder (70° tilted) for EBSD analysis. To get high quality EBSPs, the EBSD data were acquired at an acquisition speed of 20.76 Hz, beam current of 10 nA, resolution of 622 × 512 pixels (sensitivity mode) and exposure time of 6 ms/pixel. The acquired Hough-transformed orientation data from the tri-crystal Si sample was refined through pattern matching with dynamical patterns (as per (Winkelmann *et al.*, 2021)) within *MapSweeper* in AZtecCrystal v3.3 (Oxford Instruments, UK).

For ECP analysis, two sets of reference patterns were generated: (a) using EMsoft v5.0, to generate reference spherical patterns that could be reprojected to any particular orientation; and (b) using the same simulation engine as *MapSweeper* but controlled by a custom front end script, which is the so-called Bloch Wave Kikuchi Diffraction (BWKD) approach (Winkelmann *et al.*, 2021) to generate singular projections with very high angular resolution ECPs. For both simulation approaches, the minimum $d_{hkl}$ threshold of 0.2 Å was set to include more reflectors in the high-definition dynamical simulations that were used for ECCI analysis. *MATLAB* R2023a was used for building *AstroECP* for the implementation of pattern reprojection functions within AstroEBSD, as well as for regression and cross-correlation analyses.

For SA-ECP and ECCI of surface threading dislocations in GaAs over Ge crystal, a 5000 × 5000 pixel high resolution dynamical Kikuchi stereogram was simulated at 20 keV using *EMsoft* and the patterns were reprojected within *AstroECP*. For this sample, ECCI micrographs were acquired at electron beam energy of 20 keV and beam current of 998 pA at a working distance of 7.5 mm. For storing high quality micrographs, the acquisition speed was kept at relatively slower rate of 1 ms/pixel (i.e. 262 s per frame), with a pixel size of 3.91 nm and a spot size of 4.12 nm.

### 2.3. Overview of the experimental setup for SA-ECP and ECCI study

Importantly for doing ECCI work where the illumination conditions are well known, the SEM should be able to form an ECP, which requires rocking of the incident beam through relatively large angles, in the selected area approach about a small area (or ideally a point). This can be achieved using different deflections of the beam within the column, and one method available within TESCAN instruments is the 'Channeling' mode. This configuration is schematically shown in Figure 1a, where the electron beam is systematically 'rocked' about a selected area to form a micrograph where the (x,y) points of the micrograph image correspond to a projection of the rocking pattern, such as the gnomonic projection as used here, to provide contrast that is associated with the change in incident beam vector [uvw] with regard to the illuminated selected area. For SA-ECP formation, the electron lenses and scan coils are used to rock the beam and change the angle at which the beam interacts with the crystal lattice, and in practice this produces Kikuchi patterns that contain distinct intersecting bands that result from the channeling of the electron beam by specific lattice planes and zone axes of the crystalline structure. These channeling-in ECPs are akin to the channeling-out Kikuchi diffraction patterns, as found in the more commonly used EBSD technique.

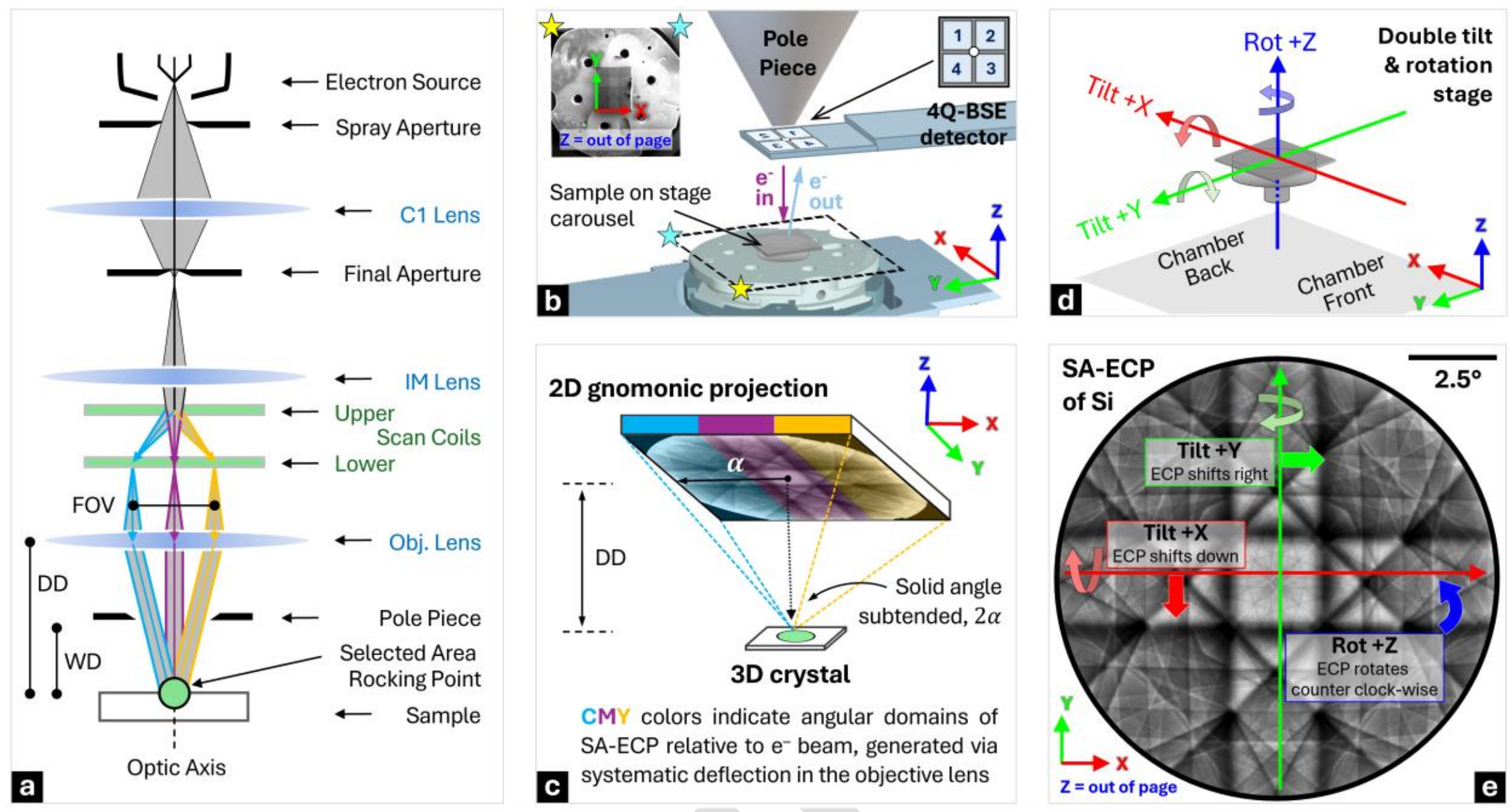

**Figure 1** (a) Schematic ray diagram for acquiring SA-ECPs in *Channeling mode* in a TESCAN AMBER-X. The central point in pattern lies on the optic axis of microscope, where the beam is 'un-rocked', is termed as the direct beam (b) A 3D model view inside the SEM chamber. Inset features a 4Q-BSE image of a Si sample showing channeling pattern. The backside of the chamber is associated with right edge of the micrograph. (c) Schematic of generation of 2D gnomonic projection from a 3D crystal sample, subtending a solid angle (2α), projected at a distance that is known as detector distance (DD) (d) Schematic depiction of a stub-mounted specimen and its +/– sense of tilt and rotation directions (e) Axis conventions and how the changes in stage configurations affects the channeling pattern. This experimental SA-ECP is captured in *Channeling* mode from a Si [001] single crystal featuring a <100> zone axis at the center.

To achieve high contrast in a SEM, typically the channeling related signal is recorded by measuring the total backscatter electron signal via a large angle semiconductor backscatter (or forward-scatter) detector. In this work, we use a 4Q-BSE detector that is placed under the pole piece (Figure 1b). However, we note that all the incoming electrons can channel when they strike a crystalline sample, and this can result in a modulation of the resultant signal collected (e.g. we routinely also collect

ECPs when examining the concurrently collected SA-ECP from the Everhart–Thornley (ET) detector, which usually is optimized to collect secondary electrons).

In the SA-ECP 2D gnomonic projection such as that shown in Figure 1c, the (virtual) detector distance (DD) directly influences the solid angle[3] subtended (2α) and this is related to the distance of the selected area (i.e. pivot point) from the objective lens. In practice, a shorter DD captures a wider angular range of backscattered electrons, resulting in larger scattering angles being projected and greater field coverage in the pattern, which manifests visually as a decrease in apparent magnification – a "zoomed-out" effect. Conversely, a longer DD results in smaller subtended angle across the pattern, but importantly a smaller angle subtended per pixel within the ECP (i.e. a higher angular resolution in the ECP). In a TEM, the detector distance is equivalent to the camera length, but as the *AstroECP* software is built upon the *AstroEBSD* system, we use the EBSD-based geometry conventions, where the pattern center is described as the center of the ECP and the detector distance, i.e. PC = [0.5,0.5, DD] using conventions from Britton et al. (Britton *et al.*, 2016).

Once the SA-ECP is captured, the microscope can be returned to a more regular imaging mode. If the microscope is well aligned, then the optic axis of the microscope should coincide with the central point within the SA-ECP and so the contrast in the regular imaging mode can be related to the orientation of the sample with respect to the optic axis of the microscope, i.e. the path of direct 'un-rocked' beam. If the contrast is optimized, this BSE image now becomes an electron channeling contrast imaging (ECCI) micrograph where the contrast is related to changes in the intensity of channeling-in of the beam with position across the sample. For purposes of understanding the contrast of the ECCI micrograph, at reasonable magnification, the point where the direct beam strikes the sample is usually near the central point of the SA-ECP.

In addition, the electron beam must be optimized for an ECCI experiment. To maximize the sharpness of the SA-ECP, ideally a highly focused near-parallel electron beam is required, so that there is an optimal sharpness of the features within the SA-ECP and ECCI micrograph. Contrast can also be optimized through use of a large and highly sensitive detector that is well placed to measure the majority of backscattered electrons (and suppress variations in contrast associated with small sample tilts) (Berger, 2000).

In practice, to align the sample for a particular channeling condition, the sample can be tilted/rotated using the stage to present the direct beam along a particular channeling direction [uvw], as identified precisely using the SA-ECP. For the AMBER-X pFIB-SEM used in the present work, it is equipped with a so-called "rocking stage" as shown in Figure 1d-e, which has perpendicular X and Y tilt axes, combined with a rotation axis that rotates the sample about surface normal (i.e. three degrees of freedom).

[3] Note that we opt to use the total angle subtended to describe the electron channeling pattern, as this is useful when considering angles between planes and directions in the crystal unit cell. It is also possible to calculate this in terms of the total solid angle of the projected cone, which may be more common to describe the angle subtended by detectors, such as those used in energy dispersive X-ray spectroscopy.

For an ECCI experiment, if we also include *scan rotation,* this double tilt + rotation setup provides a total of four degrees of freedom in how the sample can be aligned to change the channeling-in direction of the direct beam and the orientation of the micrograph on the screen, and this requires careful understanding of the frames of reference associated with each tilt and rotation operation. Once these frames are understood, then a user can navigate this space to access a wide variety of ECCI-based imaging conditions and analyse the associated variation in contrast systematically in their samples, e.g. with a goal of varying the contrast around dislocations and understanding the character of these defects.

### 3. Establishing microscope parameters and degrees of freedom

#### 3.1. Electron beam energy validation for dynamical simulation:

Understanding the channeling conditions for ECCI, as well as comparison of the ECPs against high fidelity simulations, requires knowledge of the appropriate electron beam and microscope parameters. One significant parameter is the electron beam voltage that determines the energy of the electrons that contribute to the electron channeling pattern (and dominate contrast associated with the related ECCI imaging conditions) (Han *et al.*, 2020; Wilkinson & Hirsch, 1997). The energy of these electrons is related to the associated wavelength that is used to predict the position of band edges and the intersection of these edges (akin to a two beam imaging condition), which can be approximated using Bragg's law, $n\lambda = 2d_{hkl}\sin(\theta)$, as well as higher fidelity simulations based upon dynamical theory where the energy is used to evaluate the scattering energy of the associated Bloch waves (Winkelmann, 2009).

In practice, a change in ECP-forming electron beam energy changes the location of band edges, intersections of bands, and the diameter of Higher Order Laue Zone (HOLZ) rings within the ECP. As we will rely on the variation in the Kikuchi pattern due to the presence of defects, understanding the energy of the electrons that form the ECP is very important.

The electron source on the AMBER-X is a Schottky field emitter and has a specification of within only ±0.7 eV of the requested beam energy (in this case, we are using 20 keV). In the microscope, the electrons are deflected through the electron column and strike the sample, where each electron can interact and may lose energy either via plasmon interactions to create a sharp peak of electrons near the primary beam energy or a more gradual energy of electrons that could be associated with slowing down due to electron scattering. These interactions can be simply modelled either via the continuous slowing down approximation (CSDA) or more accurately via Monte Carlo (MC) simulations that include discrete inelastic loss processes, such as differential inverse inelastic mean free path (DIIMFP) methods (Reimer, 1998; Winkelmann *et al.*, 2019; Werner, 2001, 2010; Dapor, 2023).

In our experiments, we noticed that the high spatial frequency features inside the Si <100> zone axis are highly sensitive to subtle changes in beam energy (Figure 2 and supplementary Figure S1), thus we can use this small region (subtending about 2° within the pattern) to determine the suitable energy that matches best with the simulation. These high spatial frequency features are part of the higher order Laue zone (HOLZ) lines that can be observed within the zone axis. This follows a similar observation about HOLZ lines that has been reported for Al [111] ECPs (Picard *et al.*, 2014), where the increase in accelerating voltage from 10 to 18 kV altered the geometry of shapes in [111] zone axis significantly. For a more elaborate discussion on the energy driven transition of fine structures inside the zone axis, the reader is referred to a recent study from Vo et. al. (Vo *et al.*, 2025). In this present work, SA-ECP were compared between simulation and experiment, numerically, using an approach similar to measurement of energy-dependent changes in the EBSD pattern (Winkelmann *et al.*, 2019) and analogous to the use of convergent beam electron diffraction (CBED) in transmission electron microscopy for precise lattice parameter determination (Tanaka, 1997; Humphreys *et al.*, 1988).

To find the best energy to represent the experimental 20 keV SA-ECP, a series of high-resolution dynamical simulations was generated using *EMsoft* and also with *BWKD* approach, while keeping the minimum $d_{hkl}$ threshold of 0.2 Å for both. The two methods were used as there are subtle variations in the approximations within the implementation of these two simulation packages. In both simulations tools, the energy of the incident electron beam was systematically varied between 18.5 to 21.5 keV in 0.5 keV increments, and the same pattern center and crystal orientation were used throughout (these parameters were found via an optimisation routine based on the *MATLAB fmincon* function using interior-point algorithm, with multiple energy simulations of *EMSoft* and *BWKD*). Note that *EMsoft* generates energy-weighted dynamical EBSD reference patterns by integrating aspects of dynamic electron-scattering theory with MC simulations to account for the energy and depth distributions of backscattered electrons (Callahan & De Graef, 2013). However, the CSDA in the MC module is known to underestimate the highest energy bin for the chosen accelerating voltage and care should be taken while choosing the energy cut-off values in the .nml files for MC and EBSD in *EMsoft*. To use the upper energy bin for each nominal beam energy in our experiment (EkeV = 18.5, 19.0, …, 21.5), we run the MC simulation in *EMsoft* at that full EkeV and then used only the highest energy bin [Ehistmin = (EkeV − 0.5) to EkeV] in the dynamical diffraction calculation for generating the associated reference stereogram. For each simulated energy, the normalized cross-correlation factor (XCF) between the experimental and simulated patterns was computed using *MATLAB's* corr2 function. The area of the SA-ECP was restricted to a region within about 2° inside the zone axis to focus the normalized cross correlation analysis on the HOLZ rings and fine zone axis structure and reduce the impact of other ECP variations and results of this analysis are shown in Figure 2. The resulting plot of normalized XCF versus simulation energy has a peak that corresponds to the energy at which the simulated pattern most closely matches the experimental data and revealed a distinct maximum at 20 keV, with a full width half maximum (FWHM) of 1.18 keV and 1.22 keV for *BWKD* and *EMsoft* simulations, respectively.

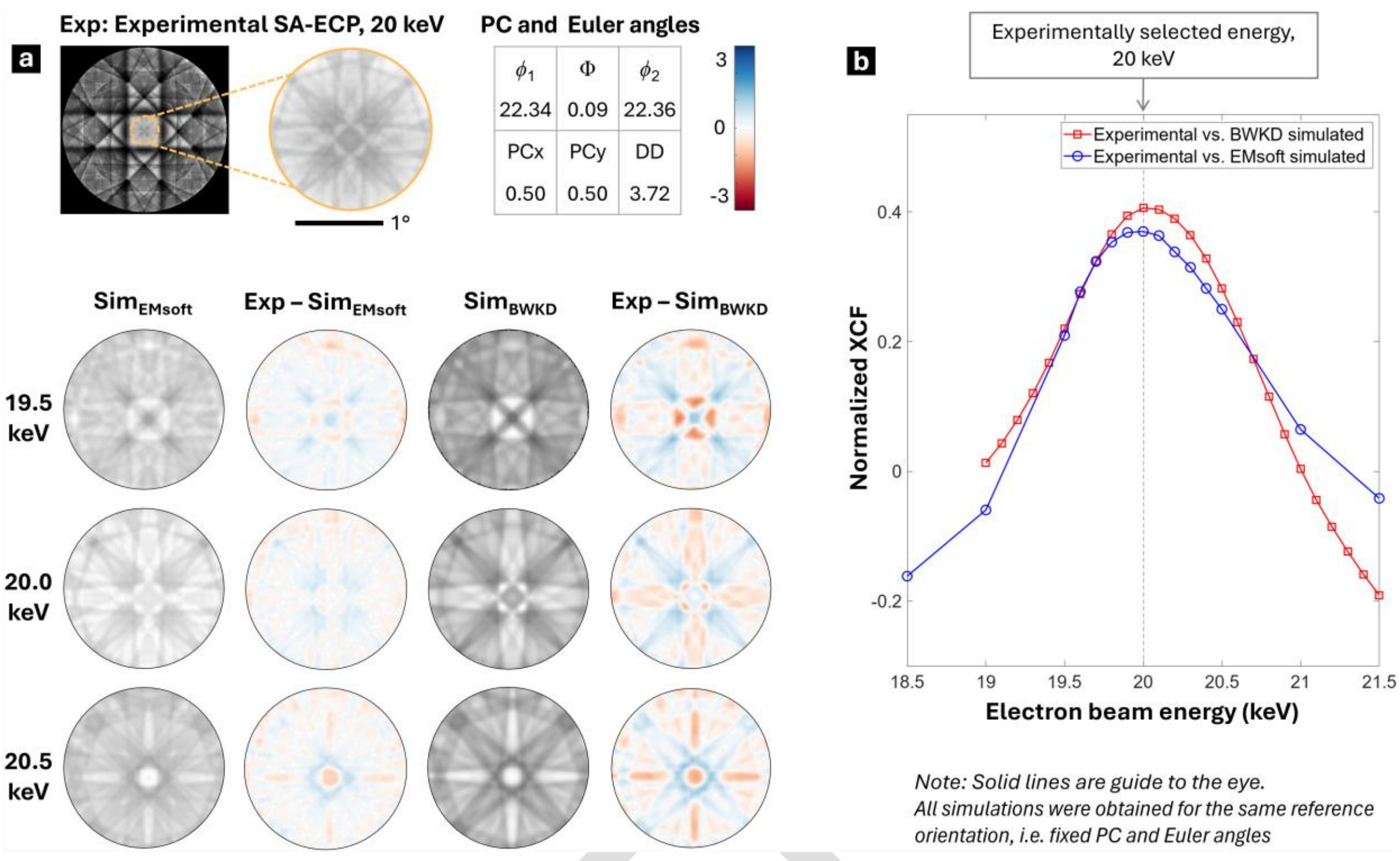

**Figure 2** Cross correlation of experimental SA-ECP (Exp) with dynamical electron diffraction simulated patterns obtained using *BWKD* ($Sim_{BWKD}$) and *EMsoft* ($Sim_{EMsoft}$); (a) A small central region of the SA-ECP was used for comparison and difference plots as it carries high spatial frequency information as shown in the inset. The similar region is shown at 19.5, 20 and 20.5 keV energies, which shows that the diffraction geometry containing HOLZ lines is highly sensitive to beam energy; (b) The energy yielding the highest normalized cross-correlation factor is the same as that for experimental SA-ECP.

### 3.2. Optimum probe current selection

Apart from beam voltage discussed in the previous section, the electron beam current is another significant parameter that may affect the beam convergence angle, electron beam spot size, and the signal to noise ratio (SNR) for an ECCI and ECP experiment. To quantify the effects of beam current on these parameters, the beam voltage and working distance were fixed at 20 kV and 7 mm respectively, and beam current was varied from 0.1 to 15 nA, with estimated beam parameters recorded from a column simulation in conventional imaging mode (ECCI) as well as channeling mode (SA-ECP), via a Python script. For the SA-ECP formation, a systematic analysis of pattern sharpness was performed by collecting ECPs from the [001] oriented signal crystal silicon sample in a constant dose experiment. The electron dose was fixed at $6.242\times10^{7}$, calculated as the product of beam current, dwell time per pixel, and the elementary charge per electron. While keeping this dose constant, the probe current was varied, and the dwell time was adjusted accordingly to examine the influence on pattern SNR and normalized XCF with the simulated pattern.

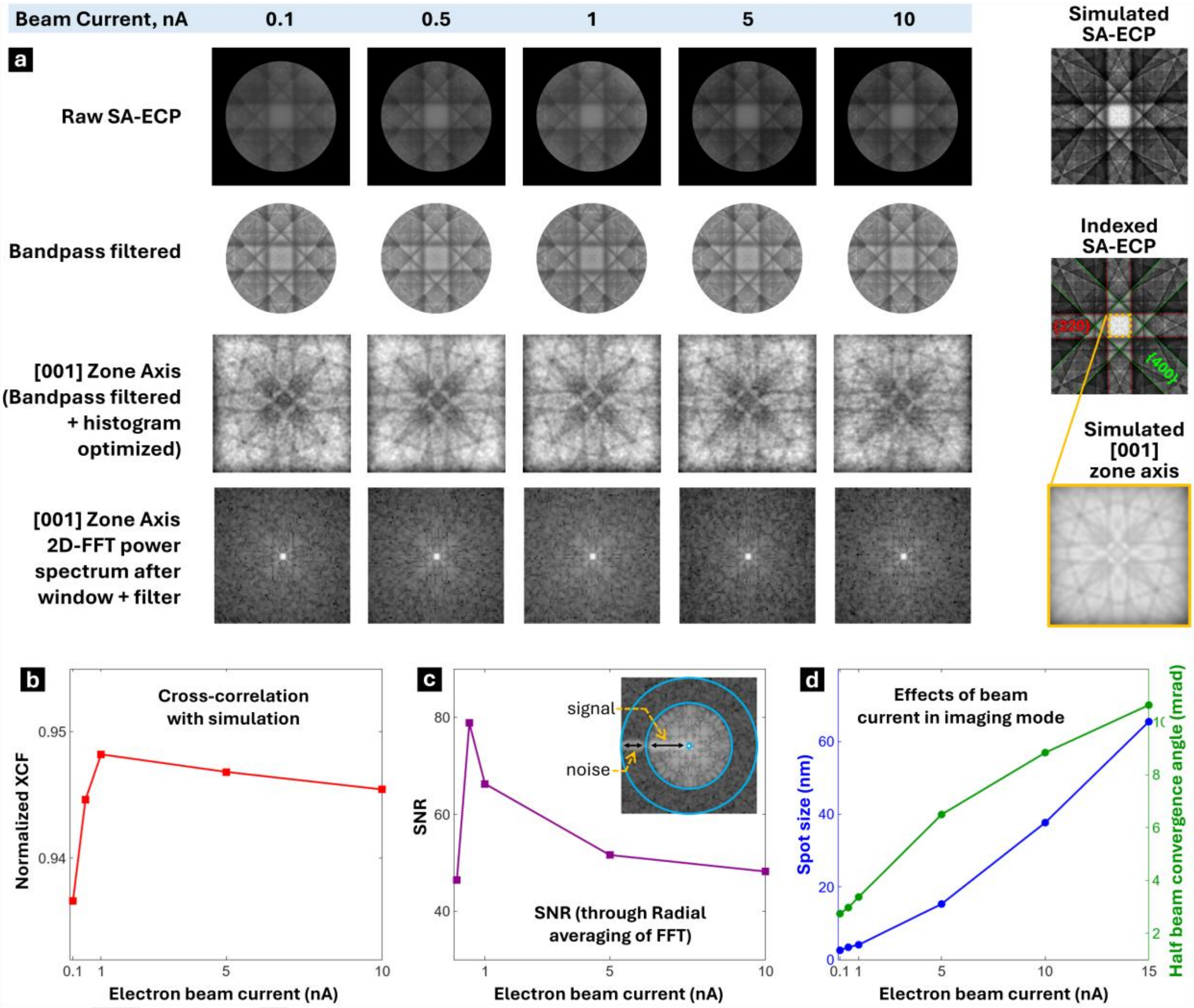

**Figure 3** Systematic collection and analysis of the beam parameters and SA-ECP with increasing probe current; (a) top row shows experimental SA-ECPs at indicated beam current that are low frequency filtered, cropped and post-processed in subsequent rows to get [001] zone axis. 2D-FFT power spectra with Hann window of this zone axis reveals the change in measurement of high frequency information; (b) correlation analysis as comparing the experimental data against the dynamically simulated ECP; (c) systematic analysis of the 2D-FFT signal to noise using radial averaging with appropriate range selection as shown; (d) increasing trend of spot size and beam convergence angle with increasing beam current in imaging mode.

The collected patterns and their post processing are shown in Figure 3, which reveals ECCI and ECP collection is best performed with probe currents of 0.5–1 nA, as this results in the highest SNR and best reproduction of the simulated pattern. Increasing the current above this range shows little or no effect on pattern appearance but significantly reduce acquisition time due to shorter required dwell times. Conversely, using lower currents will require longer exposure times, with possibility of risks such as stage drift and increased noise. Furthermore, in imaging mode increasing the beam current leads to a larger effective spot size and a higher half beam convergence angle as shown in Figure 3d and Table 1, which would in turn reduces the spatial resolution of the resulting micrographs and that is undesirable for the imaging of very fine features such as dislocations or low angle grain boundaries.

**Table 1** Electron beam parameters at various current settings

| **Beam Current (nA)** | **0.1** | **0.5** | **1** | **5** | **10** | **15** |
|---|---|---|---|---|---|---|
| **Imaging mode** | | | | | | |
| Est. spot size D50 (nm) | 2.6 | 3.4 | 4.1 | 15.3 | 37.7 | 65.5 |
| Half beam convergence angle (mrad) | 2.74 | 2.97 | 3.37 | 6.49 | 8.83 | 10.63 |
| **Channeling mode (constant dose experiment)** | | | | | | |
| Dwell time (s) | 10 | 2 | 1 | 0.2 | 0.1 | 0.667 |
| Time per frame (ms/pixel) | 2621.4 | 524.3 | 262.1 | 52.4 | 26.2 | 17.5 |
| Half beam convergence angle (mrad) | < 0.01 | < 0.01 | < 0.01 | < 0.01 | < 0.01 | < 0.01 |

Note: The spot size is calculated as D50 – i.e. the diameter where 50% of the beam intensity should be within its gaussian profile

It is worth noting that in the channeling mode, the coupling of beam convergence angle and beam current often depends on the specific architecture in terms of the column design of the SEM. For the present work, it is important to note that in the TESCAN AMBER-X system, the electron optics were designed to maintain independence of the beam convergence angle and the beam current. This is because the AMBER-X uses an independent intermediate lens (IML) to control beam focusing and convergence angle, while the beam current is regulated separately by the condenser lens. This enables the microscope to maintain a nearly parallel beam (convergence angle < 0.001° in channeling mode) even when the probe current is varied. In contrast, other SEM platforms can employ a column design where the condenser lens controls the focusing of beam from further distance to reach the narrow convergence angle and use alignment coils for scanning (Kerns *et al.*, 2020). Due to this alternative design, which may be considered advantageous for other microscopy modes, increasing the probe current inherently leads to a wider convergence angle.

### 3.3. Optic axis verification of the SEM and relationship of WD and DD:

To assess the alignment of the microscope's optic axis and quantify uncertainties in the detection geometry, a series of SA-ECP acquisitions were performed by moving the sample up and down in the chamber and therefore varying the working distance (WD) from 4 mm to 20 mm using a [001] oriented single crystal silicon sample. The captured SA-ECPs were subsequently matched with simulations using *AstroECP*, referencing high-angular-resolution dynamical simulations of silicon based on a pair of 12,000 × 12,000 pixel stereographic projections of the reference Kikuchi sphere. The pattern matching algorithm enables both the crystal orientation (i.e. Euler angles [$\phi_1$ , $\Phi$ , $\phi_2$]) and the ECP projection parameters (PCx, PCy, and DD) to be determined. For a well-aligned microscope, optic axis (i.e. PCx and PCy in the projection) is in the center of the pattern.

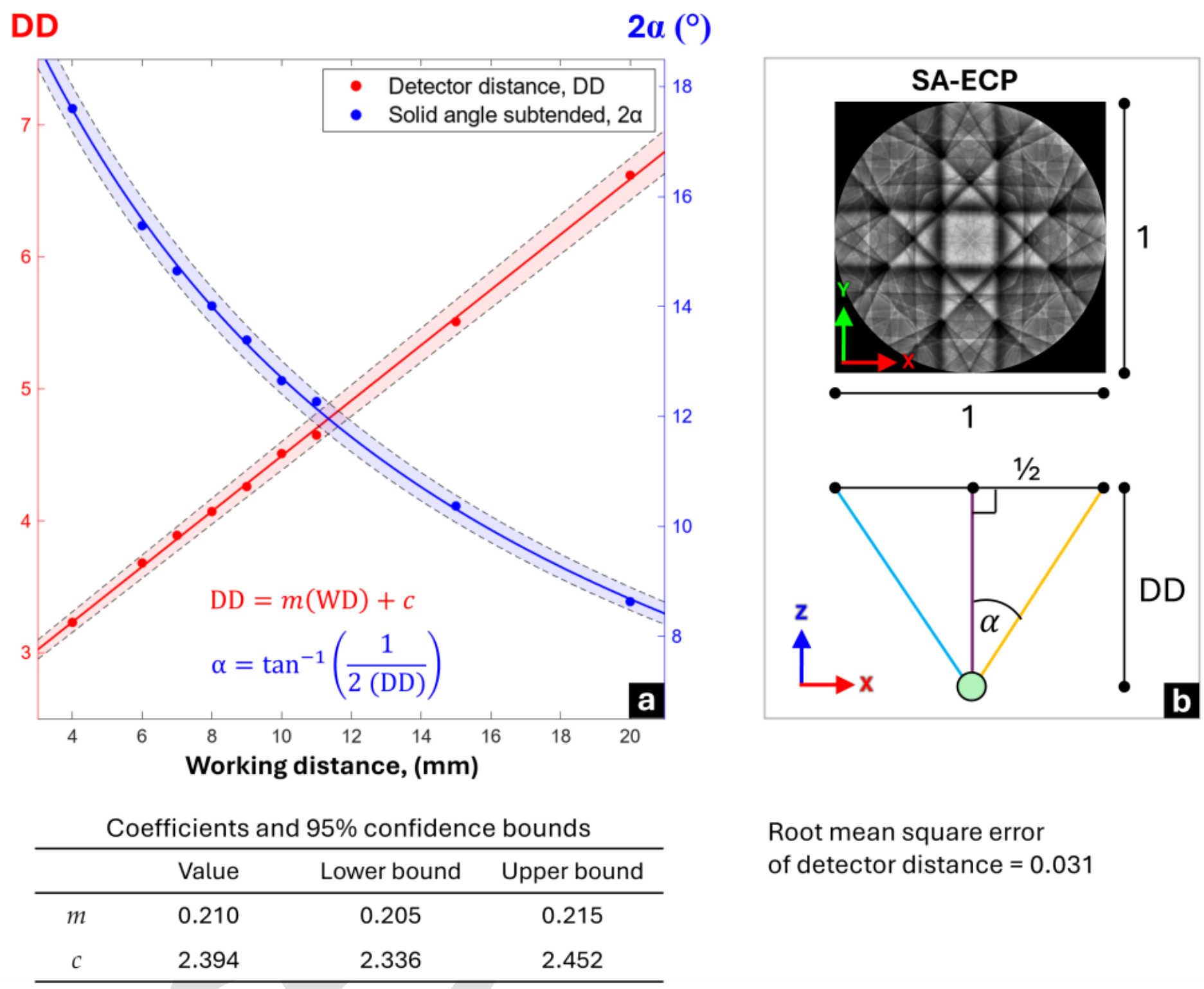

Coefficients and 95% confidence bounds

| | Value | Lower bound | Upper bound |
|---|---|---|---|
| $m$ | 0.210 | 0.205 | 0.215 |
| $c$ | 2.394 | 2.336 | 2.452 |

**Figure 4** Verification of the relationship between working distance and effective 'detector distance' for the formation of SA-ECP within the SEM (a) The measured relationship between WD, DD and 2α, where the 95% confidence bounds of the regression fit are overlaid on the plots (shaded regions) to illustrate the precision of this calibration and the coefficients are given in the table. (b) A representative experimental SA-ECP of Si with the normalized pattern width of 1×1 and a ray-tracing schematic depicting the geometrical relationship between DD, the beam origin, and the subtended angle 2α.

Figure 4 shows that as the working distance is increased, the detector distance (DD) increases linearly, and this results in narrowing of the angular field of view within the SA-ECP (i.e. a pattern zoom) that can be quantified through simple geometry (Figure 4b) as a decrease in the subtended angle (2α) across the entire pattern. This geometric relationship for the microscope working distance can be immediately used to predict the ECP subtended angle. Uncertainty of this fit is calculated through the root mean square error (RMSE) of the linear fit in *MATLAB*, and this uncertainty of 0.031 provides an indication of one of the uncertainties with regard to the position of the direct beam in the SA-ECP.

### 3.4. Stage tilt and rotation calibration and alignment checks:

Next, the stage rotation and tilt axes were symmetrically mapped: (a) to help develop a model of the stage that can be used to aid navigation of the ECP and ECCI conditions, and (b) to quantify stage-alignment errors.

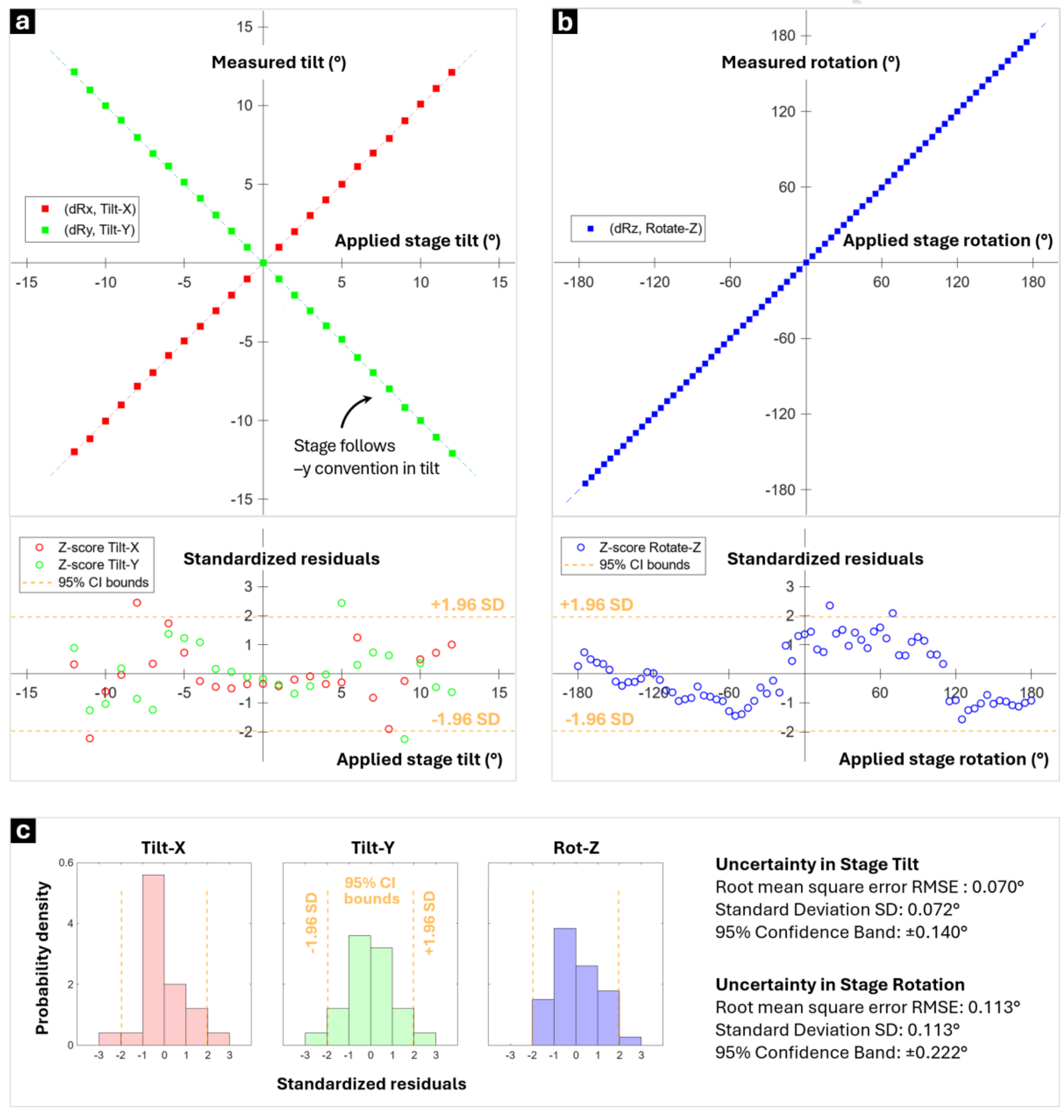

**Figure 5** SEM stage calibration through SA-ECP indexing method by measuring the indexed stage position vs. the applied position for (a) tilt about X and Y axes and (b) rotation about Z-axis. Note that our stage follows an opposite convention in Y direction i.e. TiltY = – dRy. (c) The stage uncertainty was quantified by an RMSE of 0.07° and 0.1° in tilt and rotation respectively, with standardized residuals and 95% CI bounds (±1.96 SD) confirming consistent alignment without significant outliers.

To achieve this mapping, SA-ECPs were acquired on Si [001] sample as the stage was tilted from –12° to +12° in 1° increments about X and Y axes, and also a rotation series with the sample rotated about Z from 0° to 360° in 5° steps. The SA-ECPs were indexed using the 'refine' routine (i.e. SA-ECP based pattern matching, discussed in section 4) within *AstroECP*. Subsequently, the tilt / rotation uncertainties were assessed via linear regression (via *cftool* in *MATLAB*) of the tilt and rotation series, as shown in Figure 5. Users should be aware that many stages have hysteresis, often related to mechanical backlash and other effects. Preliminary work was performed to evaluate backlash (e.g. repeat ECP collection from the same position after different stage motions, supplementary Figure S2) to ensure that these effects have limited impact on the results in this work.

## 4. Introducing GUI-based ECP analysis within *AstroECP*

It is now possible to create a 'digital twin' of our microscope within a graphical user interface to help navigate our electron channeling patterns and quickly access specific ECCI conditions for subsequent analysis, and this is achievable as we have a reasonable understanding of the microscope and stage, including: (a) the correct energy to simulate the ECP; (b) the projection parameters of the microscope; and (c) the stage tilt and rotate degrees of freedom, as detailed in the previous section.

An example of the GUI is shown in Figure 6, which shows how the GUI supports visualization of both experimental and simulated patterns side by side together with interactive navigation in 2D gnomonic space, and overlays of kinematical predicted Kikuchi bands over the patterns.

The GUI uses interpolation of a dynamical reference pattern (generated via *DynamicS*, *MapSweeper*, *BWKD* or *EMSoft*) within *MATLAB* together with calibrated controls that include the projection parameters of reference crystal orientation, and stage controls. The pattern center projection parameters use a gnomonic projection based upon the mathematics within Britton et al. (Britton *et al.*, 2016), and the optic axis is the center of the ECP, the DD is related to the working distance of microscope, as per Figure 4.

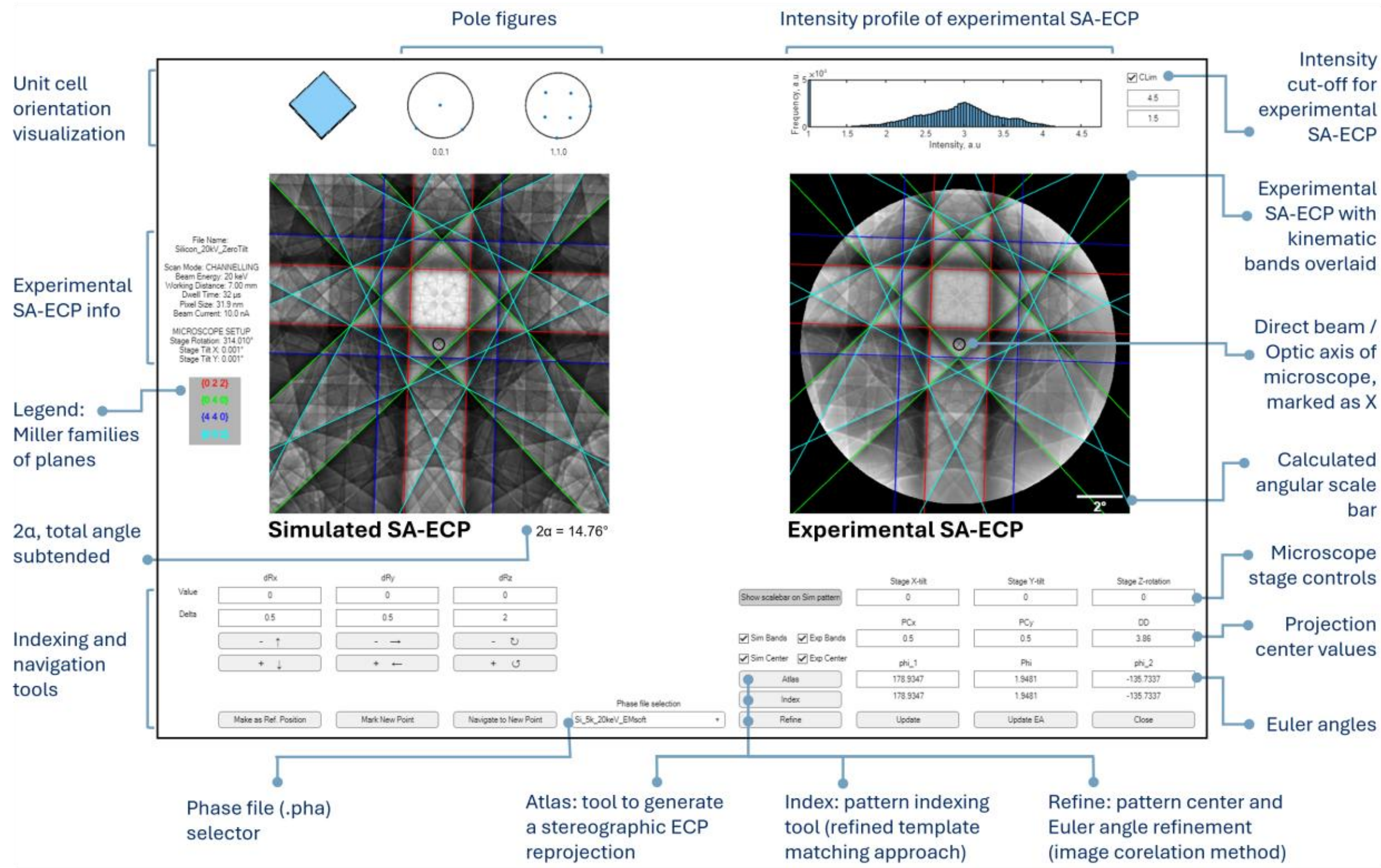

**Figure 6** A screenshot of *AstroECP*, the new *MATLAB* based GUI for indexing and navigating through electron channeling patterns. The experimental pattern is loaded as a .tif file and a corresponding phase file (.pha) is selected from a drop-down menu. The simulated pattern is then navigated through dRx/dRy/dRz buttons to match it with experimental pattern after setting the projection (i.e. projection geometry and crystal orientation). Kinematic predictions of the band edges (shown as colored lines) help in indexing & navigating the experimental pattern. The 'index', 'refine', and 'atlas' buttons are used to help match and navigate the experimental ECP.

The reference crystal orientation is described using the same pattern, which introduces the crystal orientation matrix (O) based on the Bunge convention of matrix multiplication (Britton *et al.*, 2016; Bunge, 1982)

$$O = Rz_{(\phi 2)} \times Rx_{(\Phi)} \times Rz_{(\phi 1)} \qquad \text{(Eq. 1)}$$

where ($\phi_1$ , $\Phi$ , $\phi_2$) are the Euler angles of crystal orientation describing a rotation about an axis, $Rz$ and $Rx$ are the rotation matrices corresponding to a rotation by angle $\theta$ around Z and X axes respectively.

$$Rz = \begin{bmatrix} cos\theta & sin\theta & 0 \\ -sin\theta & cos\theta & 0 \\ 0 & 0 & 1 \end{bmatrix} \qquad \text{(Eq. 2)}$$

$$Rx = \begin{bmatrix} 1 & 0 & 0 \\ 0 & cos\theta & sin\theta \\ 0 & -sin\theta & cos\theta \end{bmatrix} \qquad \text{(Eq. 3)}$$

The orientation matrix (O) can be determined initially by presenting the crystal in the microscope with no stage tilt or rotation, collecting an ECP and then matching the gnomonic projection of the simulated ECP to this reference pattern.

Next, the same matrices are used for the stage tilt (Rz) and rotation about the X axis (Rx). An additional degree of freedom is provided with the rocking stage, which allows for rotations about the Y axis, and this uses the following rotation matrix:

$$Ry = \begin{bmatrix} cos\theta & sin\theta & 0 \\ 0 & 1 & 0 \\ -sin\theta & 0 & cos\theta \end{bmatrix} \quad \text{(Eq. 4)}$$

The order of operations of the crystal orientations of the reference crystal orientation and stage tilts is important, together with the sign of the applied rotations. We note that any user can modify these within the software for other microscopes or sample stages following a similar stage mapping approach to the one performed here.

To use *AstroECP* for pattern indexing, simulation and navigation, the following setup is usually required:

- A *MATLAB* environment (R2020b or newer) with the *MTEX* toolbox (version 5.10.2 or later) and the *AstroEBSD* setup
- A representative crystallographic information file (.cif) for the sample, containing essential unit cell parameters and symmetry data necessary for accurate pattern simulation and phase definition.
- A phase file (.pha) for the material of interest, which also specifies the relevant phase reflectors for kinematic band labelling and provides the path to simulated patterns. *AstroEBSD* includes several example phase files, and users may generate new ones following the provided template for different materials.
- A dynamically simulated reference pattern generated using *BWKD* (output as .h5 files), *MapSweeper* (as .sdf5 format), *EMsoft* (also .h5 format), or Bruker *DynamicS* (.bin files).
- An experimental channeling pattern acquired from the SEM, typically in .tif format, accompanied by a text based header file (.hdr) that contains relevant microscope settings such as accelerating voltage, working distance, pattern size, stage positions etc. For instruments that do not create a text based header file, a user could create a similar file from one of the examples provided in the shared data associated with this manuscript or read metadata from the .tif file, if appropriate.

In addition to projection, initial crystal orientation, and subsequent stage tilt/rotations, the GUI also provides a simple unit cell visualization tool, as well as an overlay of the kinematic band edges to enable Miller family-based indexing of the bands within the Kikuchi pattern.

There are a few other features within this software, including:

- ECP navigation both via 'point & click' and manual adjustment through incremental tilt and rotations.
- Atlas tool: to generate a stereographic ECP reprojection of the dynamical simulation that is centered on the current projection parameters of the simulated ECP. This can be useful to guide the user if they wish to tilt the sample to a specific zone axis, or to evaluate the pattern for similar looking features (e.g. pseudosymmetry).
- Index tool: pattern matching of ECP, based upon the 'refined template matching' (RTM) approach described by Foden et al. (Foden *et al.*, 2019) and similar to ECP indexing implementation of Singh and De Graef (Singh & De Graef, 2017). In brief, the algorithm has been adapted to perform a 3D-rotation group search (known as SO(3) search) within a spacing that is ~half the subtended angle of the ECP, and then an orientation refinement is performed using the RTM algorithm. It is recommended that the detector distance is reasonably known before attempting the RTM-based orientation determination.
- Refine tool: A higher precision pattern matching based refinement of crystal orientation and/or projection parameters. This can be used once the zone axes within the experimental ECP has been identified, either from a manual match or using the RTM based search. This refinement algorithm is based on image correlation, by the interior point algorithm as implemented in the MATLAB *fmincon* function to maximize the normalized cross-correlation between simulated and experimental patterns, similar to EBSD-based pattern matching approaches (Pang *et al.*, 2020). This refinement routine is particularly helpful to provide reliable access to higher precision determination of the detector distance and crystal/stage orientation, and for example, this algorithm was used to create the graphs in Figure 4 and 5.
- Phase selection to compare different simulations or to select a different material system.
- Enhanced experimental ECP visualization (i.e. contrast stretching within the histogram).
- The GUI has also been written in such a way that features can also be used within text-based scripts, e.g. for repeat matching experiments.

The software also includes a utility to mark a "reference" crystallographic direction and navigate to a new target orientation, with the required stage motion computed and applied in the GUI to align the crystal in the microscope with a particular [uvw] along the optic axis, so that the ECCI/ECP contrast for a new zone axis or a particular band edge can be explored quickly in the SEM. In essence, this means that *AstroECP* serves not only as a pattern indexing and visualization tool but also as an experimental control assistant that facilitates ECCI by precisely guiding about the requisite stage tilts/rotations of SEM to navigate along the crystal.

We note that the calibrations (WD and DD correspondence, tilt and rotation axes) are tailored to the TESCAN AMBER-X SEM, and repeating them on other instruments may be labor-intensive, which constrains the immediate generalization of the approach. However, AstroECP can also be further developed to work in conjunction with other microscope systems, and for example a workflow for the Thermo Fisher Apreo S (PivotBeam mode) is included as an example within the GitHub release.

### 5. Demonstration 1 - Opportunities and challenges with EBSD informed ECCI

To test how well EBSD-based analysis can be used to inform ECP-based pattern analysis and ECCI-based imaging, we use the Si tri-crystal to explore whether the misorientations as measured with pattern matching based EBSD are the same as the misorientations measured with pattern matching based SA-ECP analysis. This tri-crystal was designed to easily facilitate a direct correlation between misorientations observed in SA-ECPs at zero tilt and EBSPs at 70° tilt. For this analysis, the three crystals were designated as reference (Ref), crystal 1 and crystal 2, and they are shown in Figure 7.

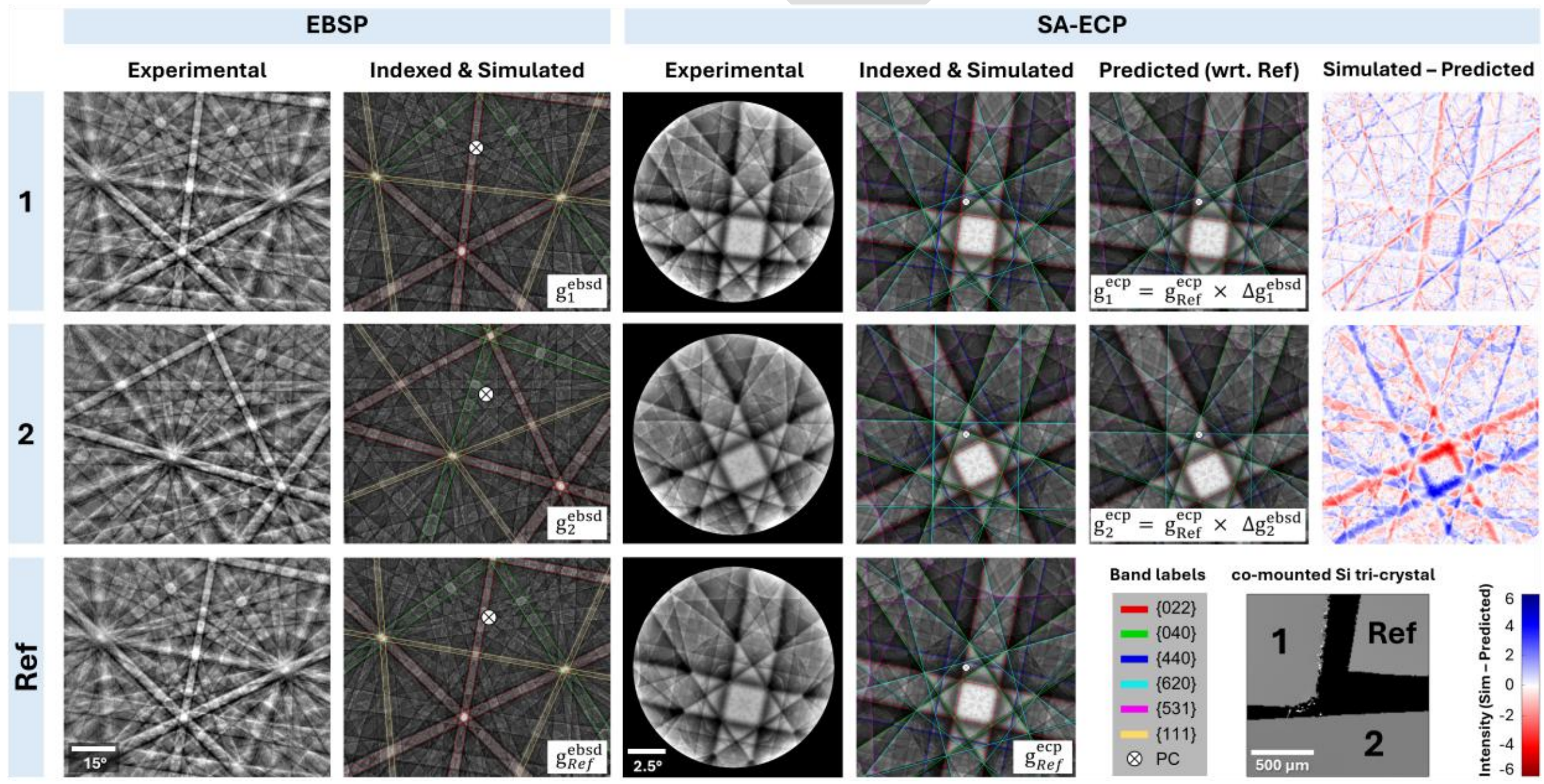

**Figure 7** Comparative analysis of misorientation correlation between ECCI and EBSD for a tri-crystal silicon sample. The six-column grid displays: (1) experimental EBSPs, (2) indexed EBSPs, (3) experimental SA-ECPs, (4) indexed SA-ECPs, (5) predicted SA-ECPs for crystals 1 and 2 with respect to the Reference, and (6) difference plots between simulated and predicted SA-ECPs. The observed discrepancies in the actual and predicted SA-ECP highlight the limitations of the EBSD-informed controlled-ECCI method for high-precision orientation alignment.

Initially, EBSD was performed on all three crystals within a single map and the acquired EBSPs were pattern matched using *MapSweeper* in AZtec *Crystal* (Oxford Instruments, UK). Within *MapSweeper*, a reasonable pattern center model was generated to describe the movement of the electron beam across the surface of these samples and the misorientations between the Reference and each of the other two crystals were calculated. The misorientation operator Δg was employed adopting Kocks et. al. (Kocks *et al.*, 2000), where Δg is described as:

$$\Delta g_x^{ebsd} = (g_{Ref}^{ebsd})^{-1} \times g_x^{ebsd} \qquad \text{(Eq. 5)}$$

where $x$ is the crystal ID, i.e. 1 or 2. Equation 5 was used to calculate an EBSD-derived misorientation of crystal 1 and 2 with respect to the reference crystal.

Following this EBSD analysis, the aluminum stub was repositioned in a standard flat configuration without tilt for ECP analysis. From a similar area within each of the three crystals, high-resolution (768 × 768 pixels, DD = 3.88) SA-ECPs were then obtained for each crystal and indexed using the refinement against dynamical patterns within AstroECP. The crystal orientations of each of the reference, crystal 1 and crystal 2 were determined and representative simulated ECPs were created. Next, using the reference crystal ECP as the input crystal orientation ($g_{Ref}^{ecp}$), simulated ECPs of crystal 1 and 2 were predicted from the EBSD-derived misorientations, $\Delta g_x^{ebsd}$, using equation 6:

$$g_x^{ecp} = g_{Ref}^{ecp} \times \Delta g_x^{ebsd} \qquad \text{(Eq. 6)}$$

Figure 7 shows additionally the comparison between the simulated ECPs derived from direct matching to the individual experimental ECPs and the predicted ECPs (that originated from Equation 6), as well as difference plots were generated to visualize discrepancies. The difference plots show that there can be large variations in intensity that demonstrate an angular discrepancy in the expected channeling conditions, which may be as large as ~0.74° as noted in crystal 2.

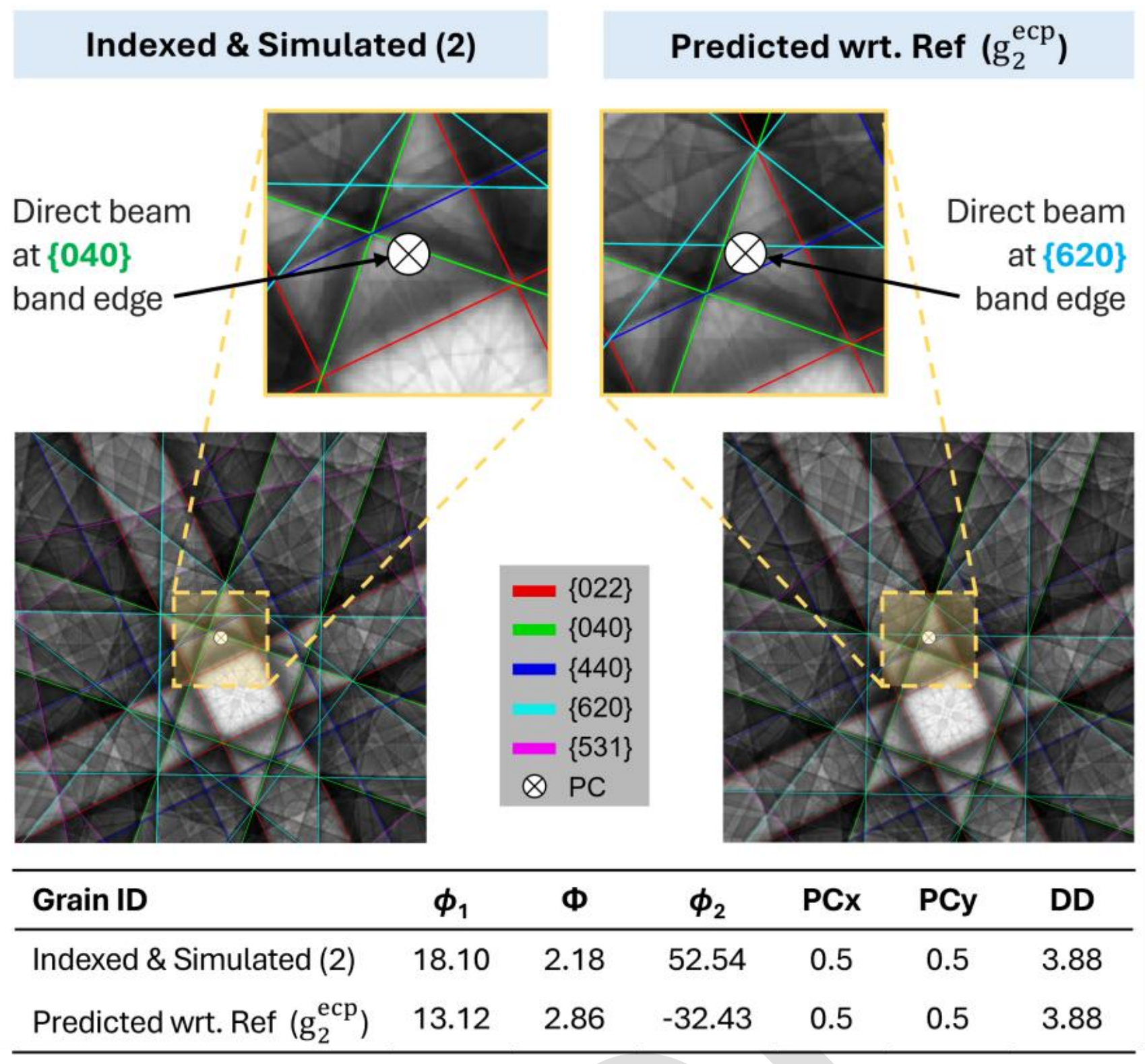

| Grain ID | $\phi_1$ | $\Phi$ | $\phi_2$ | PCx | PCy | DD |
|---|---|---|---|---|---|---|
| Indexed & Simulated (2) | 18.10 | 2.18 | 52.54 | 0.5 | 0.5 | 3.88 |
| Predicted wrt. Ref $(g_2^{ecp})$ | 13.12 | 2.86 | -32.43 | 0.5 | 0.5 | 3.88 |

**Figure 8** Comparison of simulated and predicted channeling patterns for crystal 2 as in Figure 7. Insets showing a closer look at the position of direct beam in both cases. As the crystal orientation has been shifted due to uncertainties associated when using the EBSD informed ECCI method, the direct beam lies on a different band in the predicted SA-ECP than what was in actual. The orientation information, including pattern center and Euler angles, is also tabulated for comparison.

Figure 8 shows the effect of this uncertainty in more detail, with a focus on the change in band edge that is situated along the optic axis – the experiment shows that that band edge is from the {040} plane family and the prediction would imply it is from the {620} family, which may be important for careful defect contrast interpretation using ECCI.

### 6. Demonstration 2 - Precession series for BSE contrast

The experimental validation of orientation determination was extended to a precession-assisted electron channeling contrast analysis, building upon principles from the electron channeling orientation determination (eCHORD) methodology (Lafond *et al.*, 2018). To collect the series of images in a precession experiment, the accelerating voltage was fixed at 20 kV, and a working distance of 7 mm was chosen to enhance BSE yield for our microscope. After acquiring an SA-ECP at reference position, the stage was tilted by 7° about the X-axis, so that the X-axis remained common to both the pre-tilt and post-tilt coordinate systems, schematically shown in Figure 9a. This configuration oriented the surface normal vector $\hat{n}$ at a $\sigma$ = 7°±0.1° angle relative to the incident beam vector $\hat{m}$.

Subsequently, SA-ECPs (512×512 pixel) were recorded over a precession angle ω = 0° to 360° in 5° increments, while corresponding BSE images were collected every 1° keeping a 30 µm field of view.

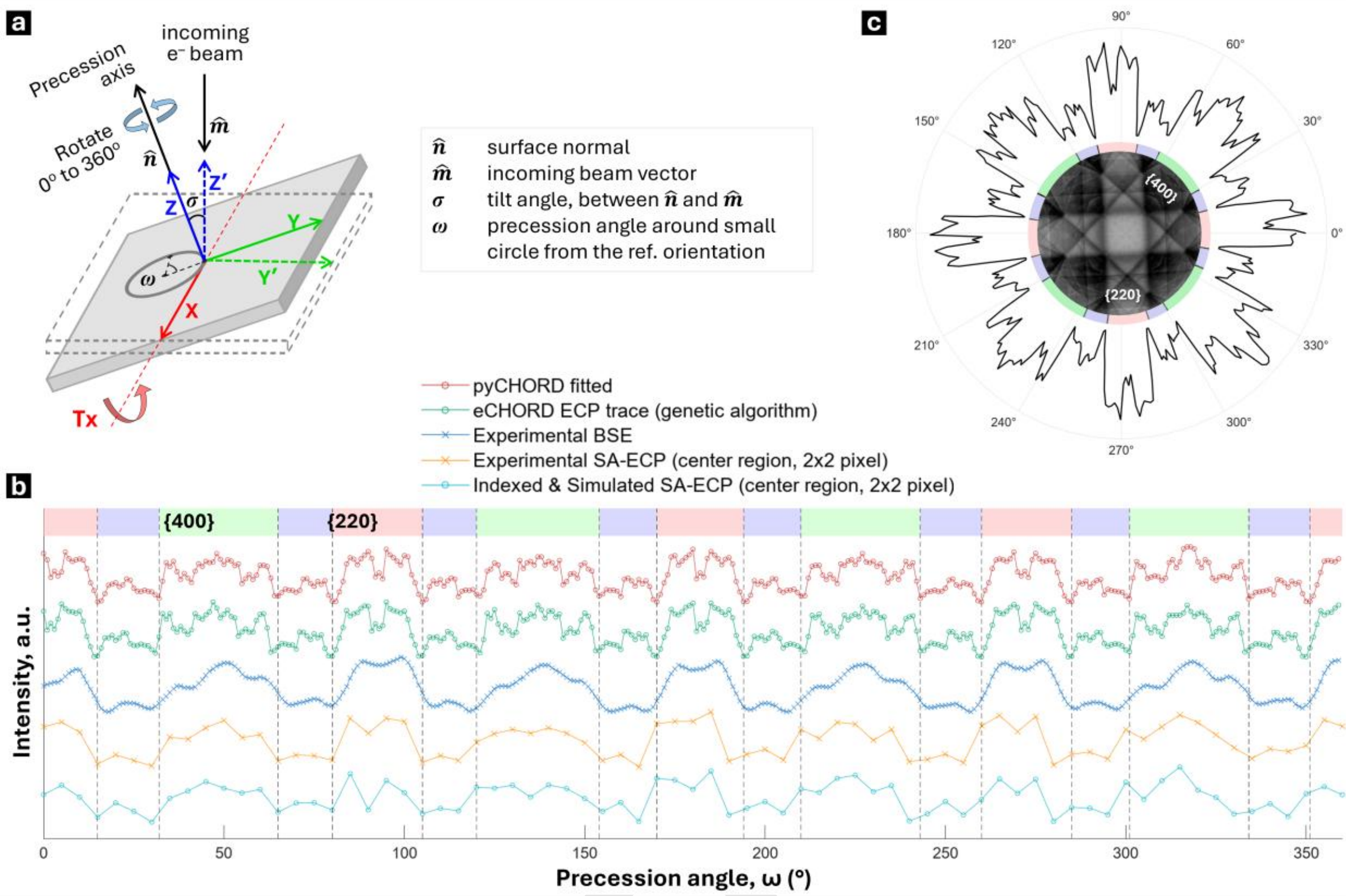

**Figure 9** Precession series analysis linking SA-ECP intensity and BSE contrast. (a) Schematic representation of the precession setup: the sample normal $\hat{n}$ is tilted by σ = 7° relative to the incoming beam direction $\hat{m}$, and precession is performed by rotating around $\hat{n}$, sweeping through ω = 0-360° while acquiring SA-ECPs and BSE images. (b) Comparison of intensity profiles extracted from the experimental SA-ECP series, dynamical simulations, BSE image series, pyCHORD orientation search, and genetic algorithm based fit, all plotted against precession angle. (c) Polar map visualization of precession intensity centered around the reference SA-ECP, highlighting the angular relationship between Kikuchi band positions (color-coded) and intensity modulations around the precession circle.

The average intensity of the four central pixels in each SA-ECP was extracted in *MATLAB* and plotted against ω to get the experimental channeling intensity profile (Fig. 9b). Simulated SA-ECP profiles were generated by dynamical reprojection in *AstroECP* and were processed in the same way as their experimental counterparts. BSE intensity profiles were computed as the mean gray level of each BSE frame vs. ω.

Two independent approaches were used to optimize the fit between the experimental BSE and simulated SA-ECP intensity profiles, i.e. a Python-based orientation search algorithm called pyCHORD (R. Scales, unpublished) and a genetic algorithm-based fitting tool in *MATLAB* built within *AstroEBSD*. For both, the objective was to fine-tune five key parameters: three Euler angles that in effect describe the precession axis, one parameter that is the radius of the precession circle linked with the tilt σ, and one final parameter that is the phase offset ($\omega_0$) around the small circle.

The search space in pyCHORD approach was sampled at 0.02° resolution without any smoothening to ECPs and assuming stage tilt between 6° to 7.4°. The resulting best-fit orientation trace is also shown in Fig. 9b.

A similar approach was followed to pyCHORD using the genetic algorithm (GA) to find an optimum parameter set, with initial estimates for the Euler angles taken from the *AstroECP* and the other parameters initialized based on experimental geometry. For the GA-approach, constraints were imposed on the search space including: the Euler angles were allowed to vary within ±1°, the precession radius within ±1°, and the phase offset within ±10°, and these bound the GA search limits. The GA-optimization aimed to minimize the difference between the experimental BSE intensity profile and the simulated profile. The GA parameters were not optimized for speed and instead were carefully tuned for high accuracy and repeatability, using a large population size of 1000, a maximum of 100,000 generations, and a strict function tolerance of $10^{-10}$ were used to search the error surface. After convergence, the best-fit solution from the GA-approach was used to reconstruct the normalized intensity profile for simulated SA-ECP series (Figure 9b).

Figure 9c presents the SA-ECP at the reference orientation, placed at 0° on a polar plot of precession intensities; the angular variation of Kikuchi band contrast (highlighted in the colored bar) mirrors the intensity modulations around the circle. Using the pyCHORD-derived center of the precession circle, the direct-beam orientation was determined to be within 0.04° of the nominal [001] axis, demonstrating <0.1° precision in optic-axis alignment (Figure 10).

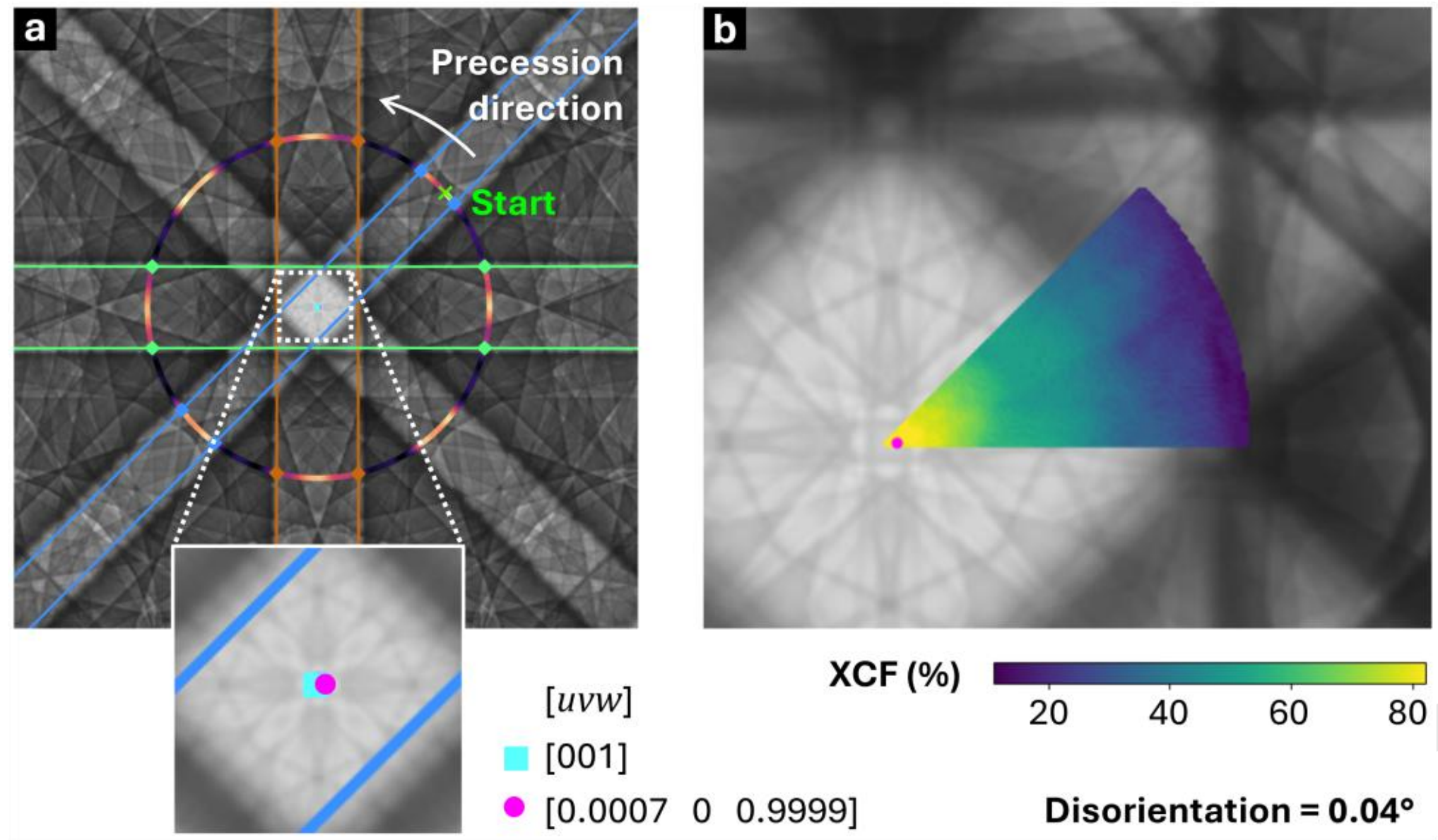

**Figure 10** Determination of uncertainty in direct beam from the invariant point (center of precession circle). (a) Simulated SA-ECP of Si showing the pyCHORD precession circle, where + sign marks the start of precession. Inset zooms in at the center of [001] zone axis, showing the center of the precession circle. The direct beam [uvw] orientation was found to deviate by only 0.04° from the nominal [001] zone axis. (b) Normalized cross correlation factor overlaid on the simulated SA-ECP for the best-fit crystal orientation determined from pyCHORD analysis of the SA-ECP rotation series

## 7. Demonstration 3 - ECCI of surface threading dislocations in GaAs thin films

An epitaxial GaAs film grown on a Ge substrate was used to illustrate the critical importance of sub-degree alignment precision in ECCI. Low dislocation density regions were identified by preliminary BSE/ECCI imaging to ensure that individual dislocations could be tracked under varying channeling conditions. A reference SA-ECP was first acquired to establish the optimum channeling condition [uvw] for defect contrast (marked position 1 in SA-ECP shown in Figure 11) near the [001] zone axis. Switching to normal SEM imaging mode at this alignment, a corresponding BSE micrograph was recorded, clearly revealing three distinct threading dislocations as regions of higher contrast (i.e. 'plumes' of black and white contrast starting from the dislocation core) that are revealed against the grey matrix, also labelled as micrograph 1 in Figure 11. To demonstrate the effect of a small angular misalignment of the direct beam on the resultant contrast for the ECCI micrographs, a systematic collection of ECCI-micrographs was performed using a small angle precession experiment as shown in Figure 11. Initially, the ECCI was collected with the direct beam directed to position 1. Next, the sample was tilted by 0.5°, an ECCI micrograph was captured (position 2), and then the sample was rotated by 45° seven further times to collect micrographs 3-9. Each of these micrographs shows a variation in the contrast associated with these dislocations, which demonstrates that a small deviation of only 0.5° from the expected [uvw] could result in uncertainty in the classification of the contrast and the associated Burgers vector of the dislocations.

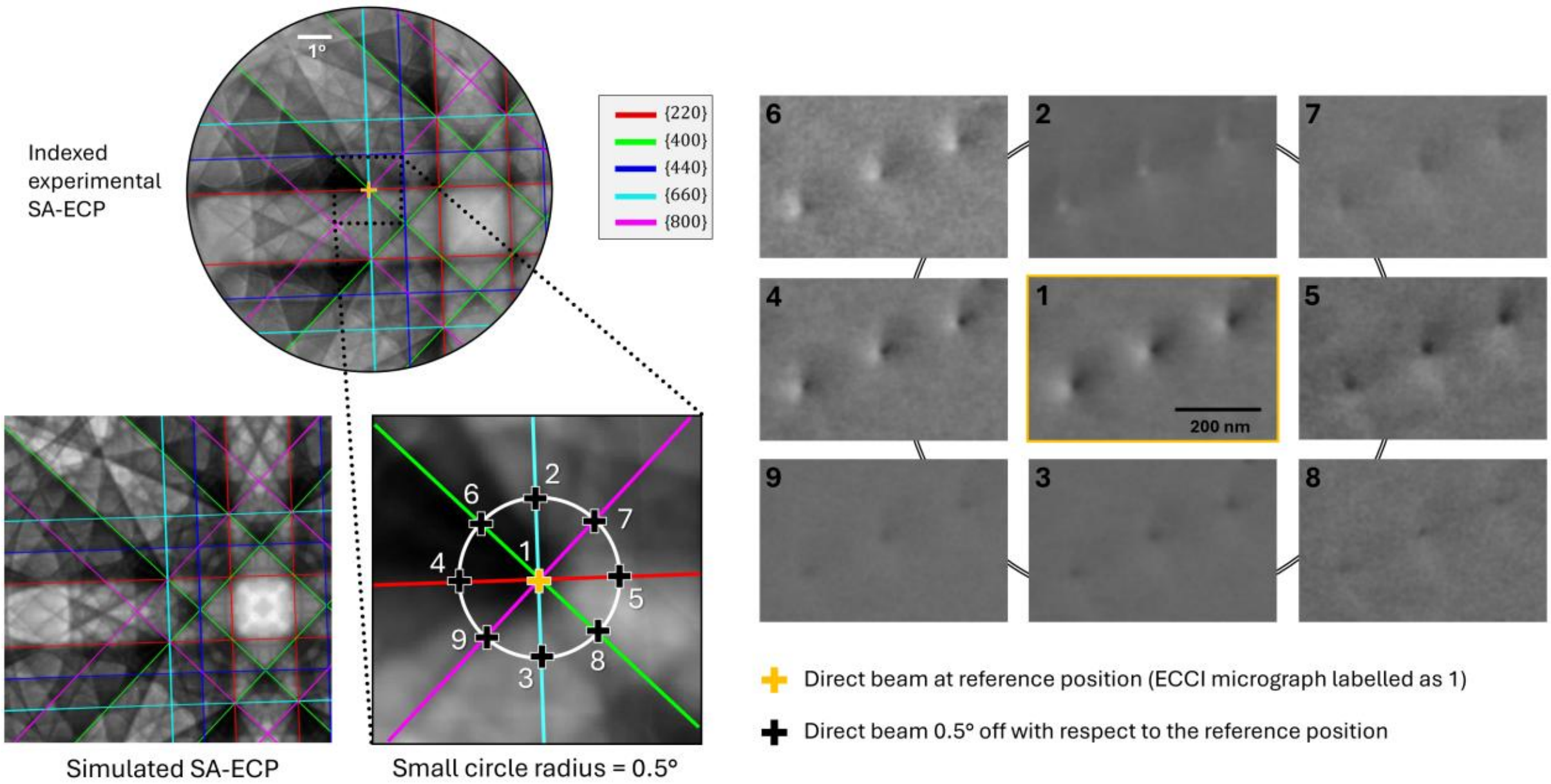

**Figure 11** ECCI of threading dislocations in GaAs/Ge. (left) Experimental and simulated SA-ECP of GaAs with the inset showing the position of direct beam 1-9 and selected diffraction conditions obtained by tilting the sample in a small circle of radius 0.5° around position 1, annotated as + symbol. (right) Mosaic of BSE ECCI micrographs acquired corresponding to channeling conditions 1-9. The micrographs 2-9 in the mosaic corresponds to a 0.5° offset in gnomonic pattern relative to position 1. The micrographs show three distinct dislocations with their contrast changing significantly with the slight change in [uvw], asserting that even a small (~0.5°) deviation from the exact channeling orientation strongly alters the observed defect contrast

## 8. Discussion

In the present work, we demonstrate systematic measurements that help understand how we can easily navigate a collected SA-ECP to optimize the incident beam direction [uvw]. This is motivated by the sensitivity of contrast in the related ECCI micrographs to different [uvw] vectors, as illustrated for the simple case of contrast around threading dislocations in GaAs (Figure 11). Additionally, we have been motivated to generate, and share, an open-source platform for ECP indexing and analysis, and to aid ECP-informed contrast optimization of ECCI micrographs. We hope this drives users to be able to increase our collective use of ECCI to aid in characterizing defects within the SEM.

It is worth mentioning that the approaches used to form SA-ECPs can differ significantly across different SEMs and manufacturers, and diverse instrumental approaches have been employed by the companies and groups, including configurations of lenses and deflection coils to achieve beam-rocking or stage-rocking for pattern acquisition (Joy *et al.*, 1982; Guyon *et al.*, 2015; Hujsak *et al.*, 2018; Kerns *et al.*, 2020; Van Essen *et al.*, 1970). These approaches influence the spatial resolution (spanning from 125 nm to several hundred micrometers), angular resolution (as low as 0.1° and over 1°) and angular range (4° to 20°), affecting the precision and area from which SA-ECP signal is generated from the sample. While analysis of these effects is part of our on-going work, optimization and determination of how these effects impact both SA-ECP formation and ECCI contrast are out of the scope of the present work.

Our calibration experiments have provided us with systematic measurements that also promote further discussion. We demonstrated that the SA-ECP geometry varies significantly with the incident beam voltage, and therefore, using the correct dynamical simulation becomes important for maintaining the angular accuracy in navigating through the crystal. Performing this analysis with both BWKD and EMsoft simulation approaches, we noticed that there are very subtle intensity variations between the two simulations, but overall, there is good agreement with the peak and the FWHM for both cases. For subsequent comparison, we suggest the use of simulated ECPs that match the energy of the primary beam energy of the microscope. This may have been implied in further studies, but our cross-correlation approach matching high spatial frequencies within the [001] zone axis provides significant additional confidence in this approach. Furthermore, one could imagine that analysis of the zone axis structure could open up more detailed crystallographic analysis, for instance building upon methods that have been developed within the convergent beam electron diffraction (CBED) community.

Prior to the experiments within this paper, we had hoped that we could use the increased precision of pattern matching between EBSD patterns to provide access to a workflow to quickly select channeling conditions, i.e. measure the misorientation between a calibrant crystal and our target region of interest with EBSD and then use this to inform our ECCI experiments. However, a rigorous comparative analysis of the correlation between EBSD and ECCI-derived misorientations has been conducted to inform best practices for high resolution ECCI based characterization workflows. The resulting difference plots (Figure 7) revealed subtle but non-negligible errors between the predicted and actual SA-ECPs, underscoring the limitations of the EBSD-informed ECCI approach and perhaps this is important for researchers who are trying to emulate the so-called 'cECCI' approach in the wider literature. Specifically, while the predicted pattern (derived from EBSD-measured misorientations via Eq. 6) approximates the general crystallographic orientation, a closer inspection of the direct beam position reveals significant discrepancy in terms of the targeted Kikuchi band edge. As we see in the related GaAs example, an uncertainty of the direct beam of this magnitude is likely very important for defect analysis because the contrast variation around the dislocation is extremely sensitive to the incident beam vector, [uvw]. This is concurrent with our knowledge in the TEM, as the $\vec{g}.\vec{b}$ invisibility criterion relies on selecting diffraction vectors ($\vec{g}$) orthogonal to the Burgers vector ($\vec{b}$).

For ECCI-based analysis, if the contrast model relies on the [uvw] vector striking a particular band edge in the ECP, it is likely important to know that perhaps a different band edge (e.g. type {620} instead of {040}) has been used for the contrast, and so the associated analysis of the contrast variation is likely to be incorrect. In reviewing the literature, this has probably been evident previously, as many studies refrain from specific characterization of an explicit Burgers vector, except in very specific crystal systems (e.g. threading dislocations in growth thin films) or classically observed dislocation types in materials with very controlled crystallography (e.g. single crystal superalloy where the orientation and dislocation types are well known *a priori,* which reduces uncertainty in slip system classification – a subtly different problem to determination). We make these comments in light of the idea that the there is an optimum contrast model that simply relies on band-edge based defect characterization, however this may not be the case and this motivates us to simply note that users should aim to optimize their understanding of the [uvw] incident beam vector with respect to the crystal (and the defects of interest). We do note that the EBSD-informed ECCI remains very useful, as the captured EBSD pattern typically subtends a very large angle (>90°) that makes it much easier to determine which zone axis is which, as compared to the relatively small angle subtended by ECCI (between 8 and 20 degrees for our set up, as shown in Figure 4). This can also help verify the ECP indexing through SO(3) based 'index' feature in *AstroECP*.

As the access to crystallographic information within the scanning electron microscope typically requires an addition of an EBSD detector, which is not available on all microscopes, the eCHORD technique has significant promise as a more general technique for grain-scale analysis. In the work here, we unite the eCHORD approach (and related methods) together with repeated SA-ECP capture, primarily to aid in our understanding of the relationship between ECCI contrast and the choice of [uvw] from within SA-ECP. Figure 9 indicates via multiple methods that, perhaps unsurprisingly, these methods converge well. Specifically, there are similar trends in the traces observed from the BSE-based ECCI precession series, as well as the central area of the ECP series, and that these can be compared favorably (via both the pyCHORD and a genetic algorithm-based search of the best matching small circle within a dynamical diffraction pattern). There are some subtle features that are of interest: (1) As a practical note, our SA-ECP micrographs have an even number of pixels so that the image center is not a central pixel, and thus we had to average over a 4 pixel central area that in effect dampens the peak and troughs and also changes the expected match in a subtle manner; (2) matching was done assuming that the precession circle was perfectly swept, and yet we know that there are some subtle variations in the rotation axis (as per Figure 5b), which could explain slight variations when carefully comparing different angular ranges within the precession circle. Despite these concerns, it is encouraging that the pyCHORD analysis showed <0.1° precision in optic-axis alignment, a level of accuracy that is reasonable for doing high-resolution ECCI where the channeling conditions are well described and as such these results support wider use of these SEM-based precession methods for high resolution ECCI studies.

The final demonstration (Figure 11) reveals the presence of threading dislocations with high quality contrast, especially when [uvw] was placed on the intersection of the {400}, {660} and {800} band edges, which is achievable for this 20 keV imaging condition. However, the contrast of these dislocations varies significantly when [uvw] is moved to differing conditions that are only 0.5° away from the central imaging condition. This echoes prior work of researchers, who discussed that tilting the sample slightly toward the inside or outside of a Kikuchi band edge respectively reduces or increases the apparent contrast of bright features, and also showed the reversal of contrast when changing the illumination conditions (Wilkinson & Hirsch, 1997; Deitz *et al.*, 2015). In TEM analysis of dislocations, where the diffraction condition can be explored through the navigation of a sparse diffraction space containing spots, the electron beam is typically slightly deviated from the exact Bragg condition by a small +/– deviation parameter ($s_g$). For example, some authors suggested that ECCI would give good contrast on the band edges where $s_g$ approaches to zero (Simkin & Crimp, 1999; Pascal *et al.*, 2018), although other disagree and reported a small positive $s_g$ for bright contrast (Kriaa *et al.*, 2017). These subtleties perhaps indicate that there is value in revisiting some of the contrast mechanisms in ECCI, and methods to approach optimal imaging conditions, and perhaps help explain some of the disagreement in the literature. Our work indicates that in practice, any residual misalignment on the order of 0.1° alters channeling contrast and can lead to misrepresentation of defect population, affecting the interpretation of contrast mechanisms that are inherently [uvw] direction-dependent. This reinforces the approach we have tried to promote in this paper, where the use of robust preliminary alignment protocols and the development of interactive pattern-navigation tools aid in the interpretation of consistent, high-resolution ECCI for consistent defect analysis.

## 9. Enhancing access to ECP/ECCI based microscopy – workflow recommendations

To aid and encourage users to develop and apply ECCI more broadly, we have used the experience gained in the formulation of this manuscript to develop a systematic and multi-tiered approach to getting started with ECP-based methods. This methodology acknowledges that successful ECCI depends not only on real-time experimental techniques but also on the foundation of prior optimization of both the microscope and the sample to make this analysis easier on more challenging samples.

Our recommended procedure (Figure 12) divides the ECP and ECCI method into key tiers that promote understanding of (1) microscope, (2) the type of sample, and (3) an individual analysis session.

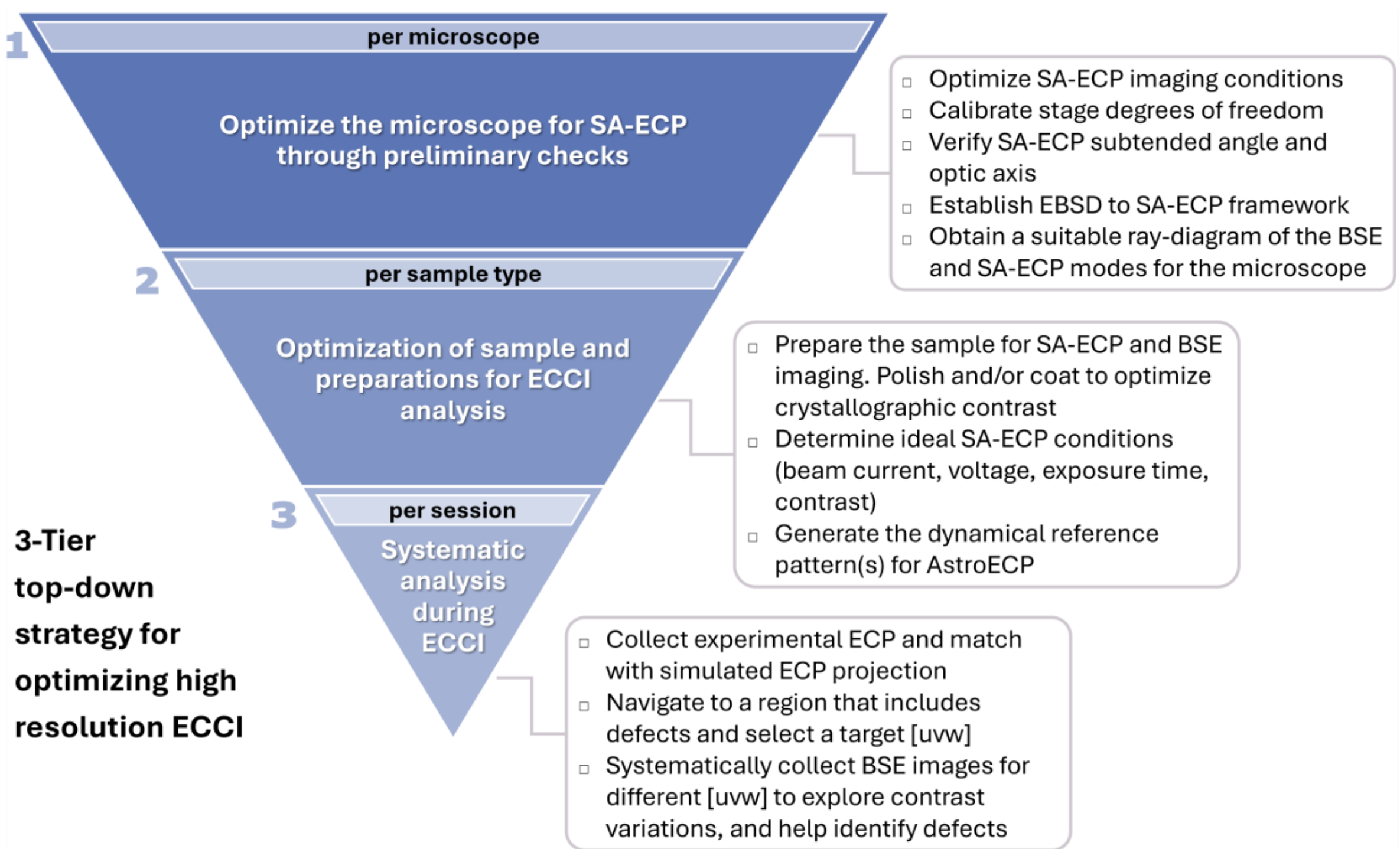

**Figure 12** Recommended approach for effectively performing ECCI, subdivided into three tiers.

Here we outline these three levels of strategy and hopefully inspire the development of a comprehensive guide to achieve reliable ECCI.

### 9.1. Tier 1: Optimization of the microscope for capture of SA-ECPs

The first tier involves an initial understanding and optimization of the individual SEM to ensure that the microscope provides consistent and high-quality patterns that can be navigated easily for reliable defect imaging. To aid in this step, it is recommended that a ray-diagram of the SEM, showing electron paths, electron optics, detector positions and geometries, be obtained to better understand the imaging geometry and beam configuration as different microscopes will realize both ECPs and ECCI-related micrographs using a different combination of lenses and detectors. To be specific here, we need to know how the SA-ECP is formed in practice and how this is related to the associated ECCI-based imaging mode.

To optimize both ECCI and SA-ECP imaging conditions, the electron beam parameters must be tuned to ensure a sharp, high signal-to-noise SA-ECP. The beam settings include an accelerating voltage of 5 to 30 kV, a beam current of 500 pA to 10 nA, and the smallest available beam convergence angle, following practices suggested in prior ECCI studies (Ng *et al.*, 1999; Mori *et al.*, 2023; Zhou *et al.*, 2024; Jiang *et al.*, 2023). For each ECP and ECCI micrograph, exposure time and detector gains should be adjusted to maximize the dynamic range without saturating the detector fully or partially, particularly when using a multi-diode detector such as 4-Quad BSE detector used in this study.

Furthermore, if each ECP or ECCI micrograph is collected over a long period, then strategies to manage drift (and contamination) must be used.

For the present work using the 4Q-BSE detector, the gains (contrast/brightness) ideally must be balanced manually and individually for all four quads individually to avoid saturation and optimize the contrast range. In general, our recommendation for acquiring a good channeling pattern would be to decrease the image brightness to <5% (depending on the type of detector used), and then gradually increasing the contrast as much as possible to maximize the crystallographic contrast achievable. In practice, use of a histogram or virtual oscilloscope can be useful for this step. It should be noted that if significant sample tilting is likely to be performed in the microscope, slight variations in contrast may occur due to the influence of the backscattered electron 'plume' on the relative electron count across the BSE detector. Consequently, contrast/brightness settings may need to be adjusted slightly during the experiment after tilting to new position (or the contrast range is reduced slightly to accommodate the maximum across different tilt angles). Furthermore, if independent control of the diode settings in a segmented BSE detector is available, the signal-to-noise ratio can be further optimized. While variation of the brightness and contrast between different imaging steps can be performed, we note that for precession-based analysis (e.g. eCHORD) and quantitative contrast analysis between images, it may be important to either calibrate the image contrast or to select a contrast that is sufficiently wide to avoid over or under saturation of the detector and resulting micrograph. A coarse rotation precession series can be helpful to adjust these settings.

Next, systematic stage tilt and rotation series experiments may be performed to check tilt/rotation sense of stage and measure the angular precision in terms of deviation from linearity. Similar to Figures 5, SA-ECPs acquired at known stage increments can be matched to indexed and simulated patterns to extract the uncertainty in all degrees of freedom of the stage. By capturing SA-ECPs at different working distances, the relationship between WD, DD and SA-ECP capture angle ($2\alpha$) may be mapped which will help in verification of optic axis alignment and also assist in understanding of the trade off in working distance and SA-ECP analysis (similar to Figure 4). These systematic experiments can be performed more easily if scripting or programmatic control of the microscope is available through an API; however, care should be taken to avoid unsupervised stage motions, which could result in damage to the microscope or the sample (especially for short sample working distances).

An optional but valuable step would be to establish EBSD ↔ SA-ECP convention by capturing SA-ECP and EBSD patterns of the same sample and linking the frames of reference. This can help specifically in the identification of the individual zone axis within the SA-ECP, especially as the angle subtended with the ECP is quite small and there may be many similar looking zone axes within the reference ECP-Sphere. To optimize this approach, configurations can be obtained via stage tilt, custom holder apparatus or novel EBSD geometries to reduce uncertainty between high-tilt vs. low-tilt setups. Here, a Si single crystal can be a useful sample as a reference pair for such an experiment due to sharp pattens, and we note that a [001] crystal typically fractures along {110} crystal plans, which provides an easy reference between the ECCI micrographs and the SA-ECP. For less symmetric crystals, care should be taken to ensure that the full orientation conventions are mapped properly (in this area, readers may find the suggestions from Britton et al. useful (Britton *et al.*, 2016)).

### 9.2. Tier 2: Optimization of sample and preparations for ECCI analysis

The second level addresses sample-specific preparations and optimizations, ensuring that the sample surface and crystallography are suitable for channeling-based imaging. Sample surface preparation should be performed to get a 'mirror polish' to suppress topographic contrast, and if required a thin conductive coating (e.g. of carbon) can be applied if charging is a concern.

For the first ECP/ECCI experiments, it is recommended to begin with an undeformed, annealed, and ideally large-grain (~100 µm) sample to acquire high contrast channeling patterns. The large grain size for this step is helpful to manage issues associated with uncertainties on the selected area being probed (e.g. due to microscope aberrations) and alignment of eucentric or compucentric tilt and rotation of the sample within the microscope. Smaller grain sizes (or more deformed samples) can be used once the user is confident with the selected area size and how to track the same region while conducting stage movements within the SEM.

Next, the user can capture a high-quality SA-ECP and determine the ideal imaging parameters (beam current, voltage/energy, pixel size, spot size, working distance) that can be standardized for subsequent ECP and BSE defect imaging. Generally, lower beam energies limit the interaction volume, providing a more surface-sensitive image with better detail on near-surface crystallographic defects, as related to beam extinction distance ($\xi_g$) (Wilkinson *et al.*, 1993; Kamaladasa & Picard, 2010). However, the BSE signal intensity is typically reduced at lower keV, which has to be compensated with higher exposure time per pixel if better image quality is needed. Furthermore, a variation in the incident beam energy can change the widths of bands in the ECP and adjust the relative positions of the direct beam direction [uvw] with respect high contrast defect imaging conditions.

For the electron probe, in most microscopes a shorter working distance offers improved spatial resolution and a wider backscatter detector collection angle allowing for a higher BSE signal and better imaging quality. However, with a shorter working distance, there is a higher risk of physical collision of the sample/stage with other components of the SEM, such as detectors or the polepiece, especially during movements of the sample on the stage. This will vary from microscope to microscope. In our microscope and detector setup, the optimum imaging conditions are typically between a WD of 4 to 7 mm (and we have two chamber scopes and a high-quality digital collision model to reduce the risk to the microscope). Conditions for different sample tilts and areas also need to be checked for achieving maximum BSE yield. As noted, tilting the sample may impact saturation of a detector signal, depending on the position and size of the BSE diodes.

After getting the experimental SA-ECP, a representative crystal information file (e.g. '.cif') should be sourced and high-resolution dynamical reference simulated spherical pattern needs to be generated. Care must be taken with simulation method, e.g. the threshold values for the strong/weak beam approximation, the selected voltage and the material parameters such as the <a>, <b> and <c> conventions and origin of the unit cells basis vectors. As established in section 3.1, many of these parameters can be optimized through comparison with an experimental silicon pattern.

### 9.3. Tier 3: Systematic experimental workflow during ECCI analysis

The third level provides the strategy to be executed during each experimental session in ECCI studies on SEM. To begin an ECCI analysis session, it is recommended that an experimental SA-ECP be captured at 'reference', un-tilted orientation and supplied to *AstroECP* (or any other pattern indexing tool) to index and match it with a dynamical simulated projection. This matching process should ensure the crystallographic orientation of the sample as identified by the Kikuchi band and zone-axes arrangements in channeling pattern. The accelerating voltage used during imaging, the working distance and the angle subtended ($2\alpha$) should be verified. Alignment of the above-mentioned parameters ensures that the beam–sample geometry is well understood and that subsequent contrast variations can be confidently interpreted in relation to the crystal lattice.

Following this, attention should be given to selecting appropriate imaging conditions for defect visualization. As our example shown in Figure 11, you can navigate to a region of interest (ROI) that contains known or suspected type of crystal defects. Next, the SA-ECP can be navigated to be coincident with potential [uvw] directions of interest, which may include low index band edges. The precise nature of contrast variation within a sample is not explored in the present work, and this will vary from sample to sample. This contrast is still subject to some discussion in the literature for a 'general' sample, even though there are very good simulations of, for example, the contrast associated with threading dislocations in $SrTiO_3$ and GaN at different imaging conditions (Pascal *et al.*, 2018; Picard *et al.*, 2014).

## 10. Conclusion

We have addressed some of the challenges that have historically limited the widespread adoption of ECCI for crystallographic defect characterization. Through systematic calibration experiments and the development of the *AstroECP* software tool, we have established a comprehensive workflow that enhances the accessibility and practicality of ECCI for wider outreach. The integration of calibrated understanding of stage tilt/rotation axes and dynamical pattern simulation enables precise navigation of SA-ECPs within quantified uncertainty metrics. Our comparative results underscore that, unlike conventional cECCI methods whose reported alignment accuracy often exceeds 0.5°, reliable high-resolution ECCI demands SA-ECP precision well below half a degree. The multi-tiered strategy we propose – encompassing microscope optimization, sample-specific calibration, and session-specific procedures – provides materials scientists with a structured pathway to implement ECCI in their research, bridging the gap between the theoretical promise of ECCI and its practical application. Future integration of machine learning for dynamical pattern recognition and automated defect classification could further amplify the technique's impact.

**Acknowledgements** The authors would like to deeply acknowledge Dr Aimo Winkelmann for the discussions and support with regard to BWKD simulation. The authors would also like to thank Dr Guangrui Xia for the samples of GaAs/Ge and Dr Marc de Graef for helpful discussion regarding EMsoft. We further acknowledge Dr Ruth Birch and Graeme Francolini for their valuable comments and suggestions on the manuscript. This research was supported in part through the computational resources and services provided by Advanced Research Computing (ARC) at University of British Columbia (UBC), Canada. Computing resources granted by RWTH Aachen University, Germany under project rwth-1308 were also used for initial EMsoft simulations.

**Funding Information** The authors thank a range of funders that supported this collaborative work: Natural Sciences and Engineering Research Council of Canada (NSERC) [Discovery grant: RGPIN-2022-04762, 'Advances in Data Driven Quantitative Materials Characterisation']; Punjab Education Endowment Fund (PEEF) 2023 supporting H. Qaiser; British Columbia Knowledge Fund (BCKDF) Canada Foundation for Innovation – Innovation Fund (CFI-IF) [#39798, 'AM+'] and [#43737, 3D-MARVIN]; the MITACS Globallink Research Award (# IT37646 and #IT28799); the German research foundation (DFG) within the Collaborative Research Centre SFB 1394 "Structural and Chemical Atomic Complexity—From Defect Phase Diagrams to Materials Properties" (Project ID 409476157) including the project group C02.

**Conflicts of interest** The authors declare no conflict of interests, with the exception of Jiří Dluhoš, who is employed as a Product Manager (FIB-SEM) by TESCAN group, manufacturer of the SEM used in this study. This affiliation did not influence the design, execution, or interpretation of the reported research in this study.

**Data availability** Raw data of EBSD/ECCI work is available via Zenodo repository (*10.5281/zenodo.15778499*). The *MATLAB* scripts including *AstroECP* v1 are available on GitHub as part of *AstroEBSD* (*https://github.com/ExpMicroMech/AstroEBSD/tree/main/modules/AstroECP*).

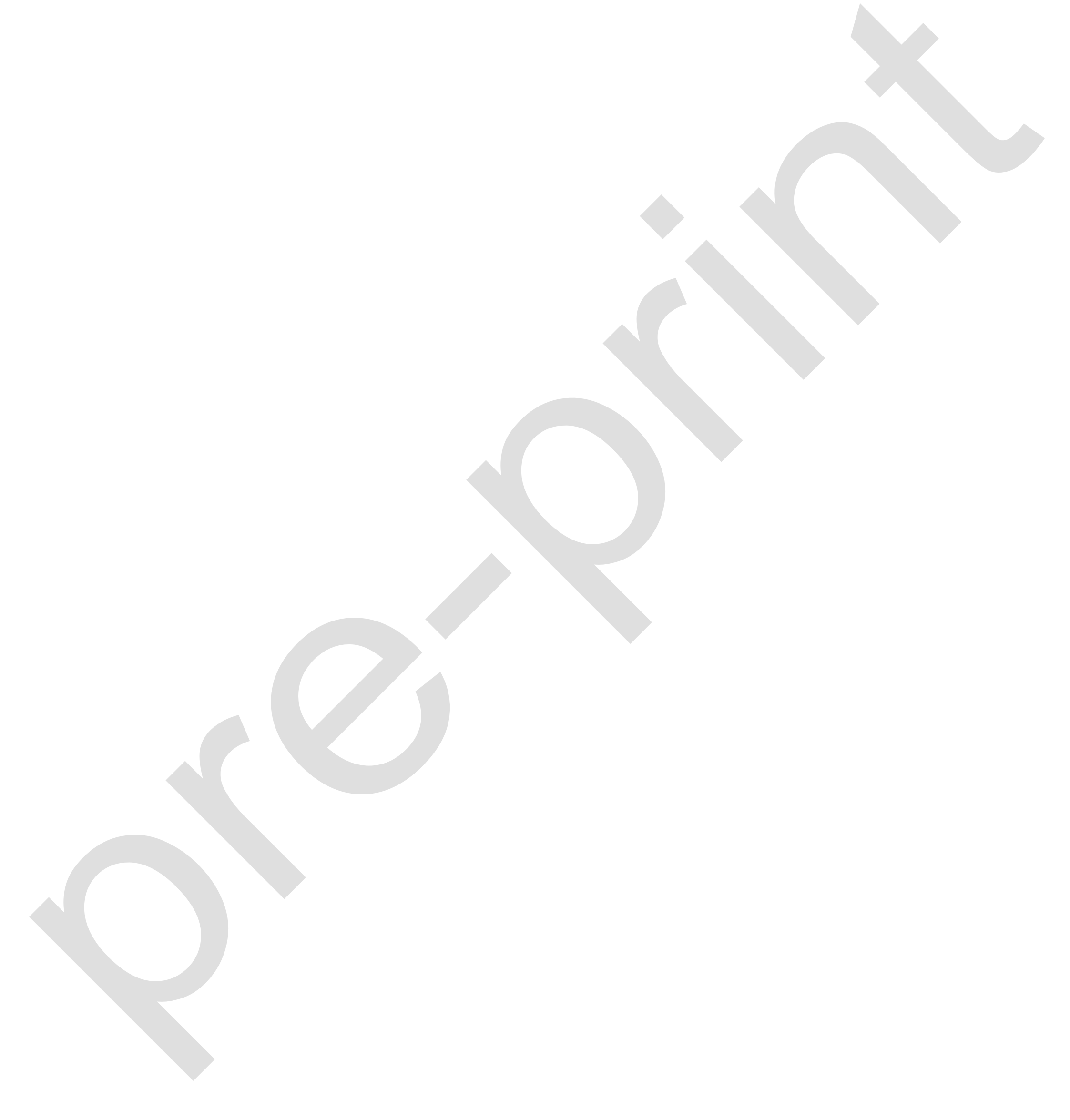

# Supporting information

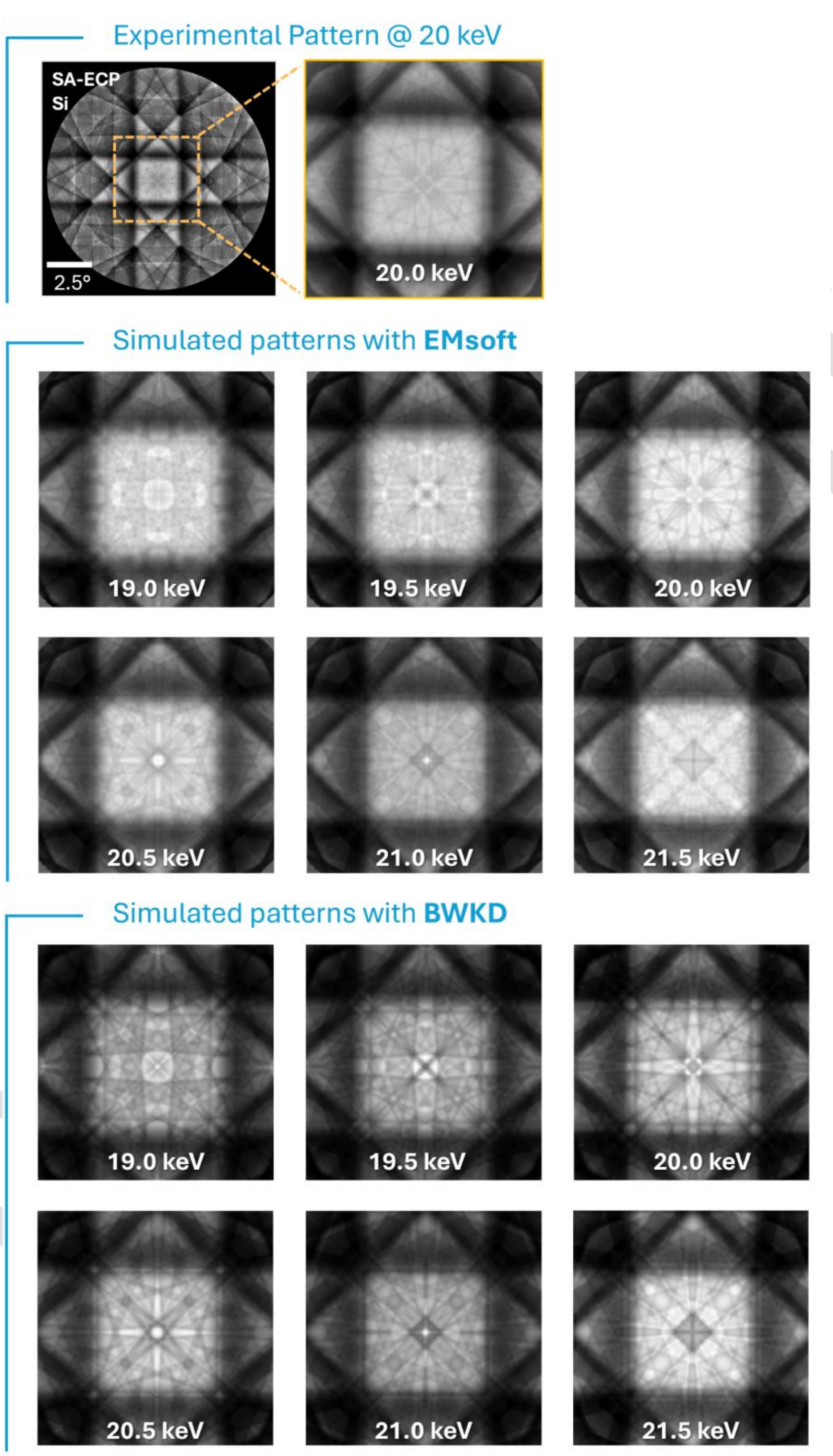

**Figure S1** Experimental vs. simulated SA-ECP at various beam energies. Note the strong dependence of voltage on the geometry of high spatial frequency features inside the zone axis

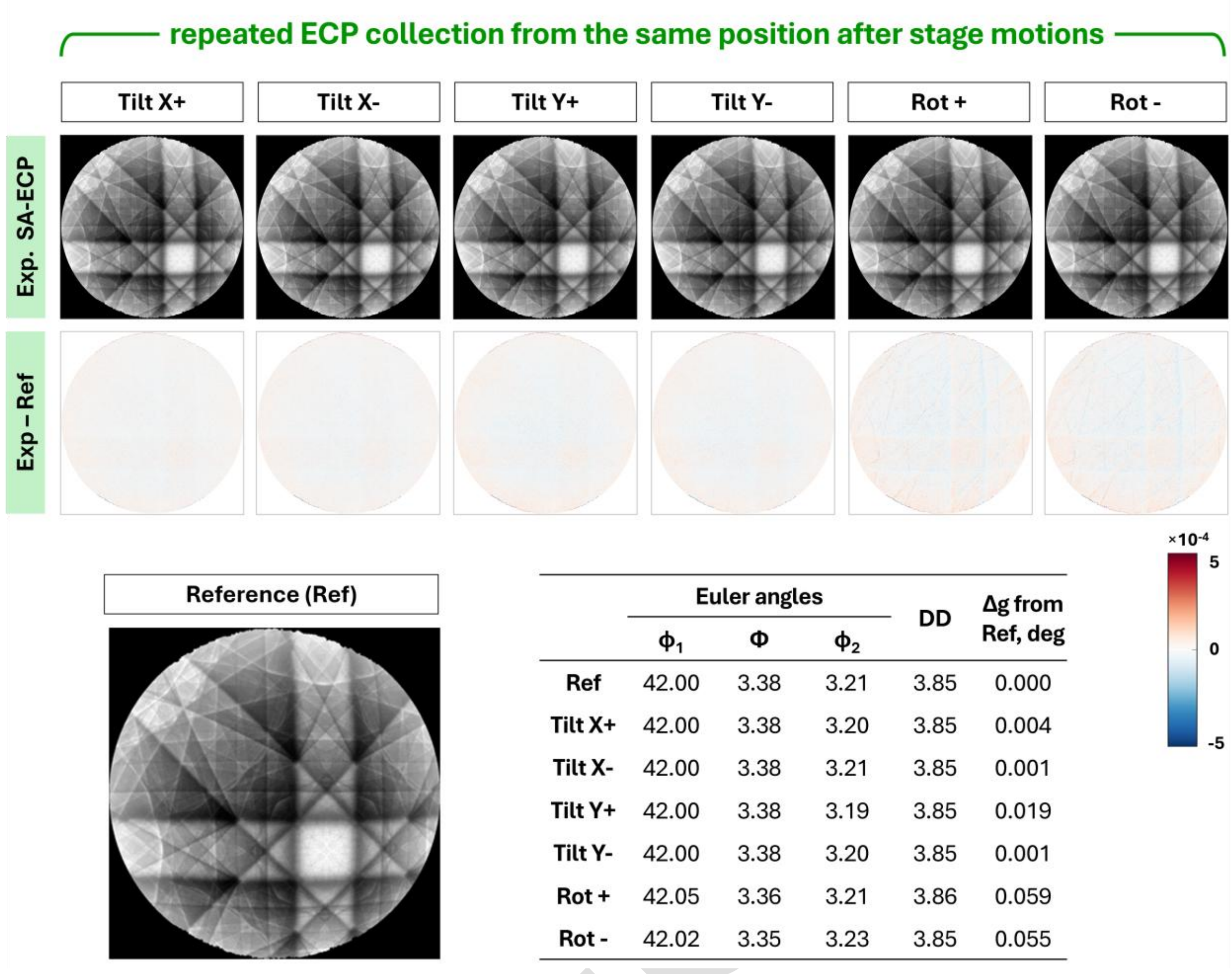

| | Euler angles | | | DD | Δg from Ref, deg |
|---|---|---|---|---|---|
| | $\phi_1$ | $\Phi$ | $\phi_2$ | | |
| Ref | 42.00 | 3.38 | 3.21 | 3.85 | 0.000 |
| Tilt X+ | 42.00 | 3.38 | 3.20 | 3.85 | 0.004 |
| Tilt X- | 42.00 | 3.38 | 3.21 | 3.85 | 0.001 |
| Tilt Y+ | 42.00 | 3.38 | 3.19 | 3.85 | 0.019 |
| Tilt Y- | 42.00 | 3.38 | 3.20 | 3.85 | 0.001 |
| Rot + | 42.05 | 3.36 | 3.21 | 3.86 | 0.059 |
| Rot - | 42.02 | 3.35 | 3.23 | 3.85 | 0.055 |

**Figure S2** Evaluation of stage backlash through repeated SA-ECP collection from the same position after different stage motions. Top row shows experimental SA-ECPs after stage motion, second row shows their intensity difference plots with respect to reference SA-ECP. The orientation data is also tabulated, with maximum misorientation Δg of 0.059° was recorded in rotation.